\documentclass[a4paper,fleqn,usenatbib]{mnras}

 \pdfoutput=1

\usepackage[T1]{fontenc}
\usepackage{ae,aecompl}

\usepackage{graphicx}	
\usepackage{amsmath}	
\usepackage{amssymb}	

\usepackage{newtxtext,newtxmath}

\title[High density PNe]{Abundance Discrepancy Factors in high density planetary nebulae}

\author[Ruiz-Escobedo \& Pe\~na]{
Francisco Ruiz-Escobedo$^{1}$\thanks{E-mail: fdruiz@astro.unam.mx},
Miriam Pe\~na$^{1}$
\\
$^{1}$Instituto de Astronom{\'\i}a, Universidad Nacional Aut\'onoma de M\'exico, Apdo. Postal 70264, Ciudad Universitaria, 04510 Ciudad de M\'exico\\
}

\date{Accepted XXX. Received YYY; in original form ZZZ}

\pubyear{2015}

\begin{document}
\label{firstpage}
\pagerange{\pageref{firstpage}--\pageref{lastpage}}
\maketitle

\begin{abstract}

From high-resolution spectra, chemical abundances from collisionally excited lines (CELs) and optical recombination lines (ORLs) have been determined for planetary nebulae Cn\,3-1, Vy\,2-2, Hu\,2-1, Vy\,1-2 and IC\,4997, which are young and dense objects. The main aim of this work is to derive their O$^{+2}$/H$^{+}$ Abundance Discrepancy Factors, ADFs, between CELs and ORLs. He, O, N, Ne, Ar, S, and Cl abundances were obtained and our values are in agreement with those previously reported. We found that Cn\,3-1, Hu\,2-1, and Vy\,1-2 have O abundances typical of disc PNe, while Vy\,2-2 and IC\,4997 are low O abundance objects ($\rm{12+log(O/H) \sim 8.2}$), which can be attributed to possible O depletion into dust grains. ADFs(O$^{+2}$) of $4.30^{+1.00}_{-1.16}$, $1.85 \pm 1.05$, $5.34^{+1.27}_{-1.08}$ and $4.87^{+4.34}_{-2.71}$ were determined for Vy\,2-2, Hu\,2-1, Vy\,1-2 and IC\,4997, respectively. The kinematics of CELs and ORLs was analysed for each case to study the possibility that different coexisting plasmas in the nebula emit them. Expansion velocities of [\ion{O}{iii}] and \ion{O}{ii} are equal within uncertainties in three PNe, providing no evidence for these lines being emitted in different zones. Exception are Hu\,2-1 and Vy\,2-2, where ORLs might be emitted in different zones than CELs. For Vy\,2-2 and IC\,4997 we found that nebular and auroral lines of the same ion (S$^+$, N$^{+}$, Ar$^{+2}$, Ar$^{+3}$, O$^{+2}$) might present different expansion velocities. Auroral lines show lower $\rm{V_{exp}}$ which might indicate that they are emitted in a denser and inner zone than the nebular ones. 
 
\end{abstract}




\section{Introduction}

Planetary nebulae (PNe) are constituted by shells of ionized gas expanding  around an evolved  hot central star which initially was a low-intermediate mass star ($1 - 8$ M$_\odot$) and it is on the way to become a white-dwarf. The shell was part of the stellar atmosphere and it was ejected in an advanced stage of evolution, during the  AGB phase. Presently the central star has an effective temperature between 30,000 K to 150,000 K and typical properties of the ionized shell are: electron densities between $10^3$ to $10^5$ cm$^{-3}$, electron temperatures of about 10,000 K, and expansion velocities V$_{\rm exp} \sim 20 - 30$ km s$^{-1}$.

The ionized shell possesses a stratified structure, with the more ionized species nearer the central star. Usually the structure shows an expansion velocity gradient, with larger velocities towards the external zones where the less ionized species are found (Hubble law flow).

A large amount of heavy ions have been detected in the nebular spectra through Collisionally Excited Lines (CELs) of N$^+$, O$^+$, S$^+$, S$^{+2}$, O$^{+2}$, Ne$^{+2}$, Ar$^{+2}$ and even lines of O$^{+3}$, Ar$^{+3}$, Ar$^{+4}$ and Ne$^{+4}$ can be detected for the cases of very high stellar effective temperature. Optical Recombination Lines (ORLs) of H$^+$, He$^+$,  He$^{+2}$ are detected in the spectrum, and faint recombination lines of heavy elements (C, O, N, Ne, ...) have been detected also.

Physical conditions and ionic abundances relative to the H$^+$ abundance can be derived from CELs as well as from ORLs. It has been found that the abundances of an ion determined from the two types of lines do not coincide, generally the abundances from recombination lines are larger than the ones derived from collisionally excited lines, giving origin to a discrepancy known as the Abundance Discrepancy Factor (ADF). Such a discrepancy has been object of numerous studies trying to understand its origin and to determine the true chemistry of the nebulae, which constitutes a fundamental parameter to understand the evolution of low-intermediate mass stars. Such ADFs are found in \ion{H}{ii} regions as well, with values 
usually between 1.5 and 3, see, e.g., \citet{liu:12}; \citet{esteban:14} and references therein, but in PNe it has a significant tail extending towards much larger values.

Several solutions have been proposed to explain the ADF problem \citep[see][for a complete  list]{espiritu:21}. 
The two most important solutions have been extensively analysed. The first one is the presence of large temperature fluctuations in a chemically homogeneous plasma \citep[and references therein]{peimbert:14}; the fluctuations are larger than those predicted by photoionized models, and they have been used mainly to analyse ADFs in \ion{H}{ii} regions through the parameter $t^2$ that represents  the size of the fluctuations \citep{esteban:14}. The second solution corresponds to bi-abundance models where small H-deficient inclusions would be mixed with the H-rich hot plasma \citep{liu:00}, these inclusions would be cold ($\rm{T_e}$ about 1,000 K) and heavy element recombination lines would originate predominantly in this plasma.  

According to \citet{peimbert:14}, who analysed more than 20 PNe, ADF values as large as 9 can be explained by the presence of thermal inhomogeneities in a chemically homogeneous plasma  without requiring the presence of the small inclusions proposed by \citet{liu:00}. However in \citet{peimbert:14} work the possibility of different plasmas of slightly different chemistry and physical conditions, coexisting in the nebulae, cannot be discarded. This possibility  can be studied through kinematic analysis and it is the one that can be explored in this work where we analyse the chemistry and the kinematics in a sample of nebulae.

Since some years ago our group have started the analysis of several PNe to determine the physical conditions and chemistry in the ionized gas by using CELs and ORLs, by means of spectrophotometric data of high- and low-spectral resolution. In our previous work \citep{pena:17} and in works by other authors, evidence of different plasmas, with different chemistry and spatial location, existing in some nebulae, has been reported: one plasma is more widespread, it has  low metallicity and high electron temperature, and it emits mainly the CELs, the other has larger heavy-element abundances and lower electron temperature, it is more concentrated towards the inner nebular zone and produces mainly the ORLs (see, e.g., \citealp{garcia-rojas:16,richer:13}).

In \citet{pena:17}, we analysed the kinematic of CELs and ORLs for a sample of 14 Galactic PNe with [WC] central star and ADFs $\leq$5, finding  that in several of them there is evidence of the existence of two different plasmas with different characteristics producing the CELs and ORLs, as mentioned above.

 Our goal in this and upcoming articles is to enlarge that work and to determine the ADFs for O$^{+2}$ of samples of PNe with different properties (central star types, low and high densities, different electron temperatures, different chemistry) to verify the validity of the hypothesis of two coexistent plasmas in PNe as a possible origin of the ADF, and to search for possible correlations of ADFs with some nebular characteristics. To do so, we analyse deep high-resolution spectra to determine physical conditions, abundances, and expansion velocities of CELs and ORLs, aiming to find if the emission of these lines arises from different regions within the nebulae. 

In this work, we present the analysis carried out for five young, very high-density nebulae, three of which have been reported in the literature to show ADFs larger than 4. The objects under study are Cn\,3-1 (PN G038.2+12.0), Vy\,2-2 (PN G045.4-02.7), Hu\,2-1 (PN G051.9-03.8), Vy\,1-2 (PN G053.3+24.0), and IC\,4997 (PN G058.3-10.9),  whose main characteristics are presented in Table \ref{tab:PNe}.  Although some of these objects have been analysed previously by other authors, it is important to redetermine the physical conditions, the chemistry and the ADFs provided by CELs and ORLs in them in a homogeneous manner, by using the same procedures to analyse our data with up-to-date atomic parameters and to determine the uncertainties involved. In addition, we analyse the kinematic behaviour of CELs and ORLs that has not been done previously for these objects. 

This paper is organised as follows. In \S2 the PN characteristics, observations and data reduction procedures are given. In \S3 the physical parameters derived from CELs and ORLs for each object, are presented. In \S 4  ionic abundances from CELs are calculated, while ionic abundances from ORLs are computed  in \S 5. Total abundances are presented in \S 6. The nebular expansion velocities, derived for the different ions from their CELs and ORLs are discussed in \S 7. The results for individual nebulae  can be found in \S 8. Our discussion and conclusions are presented in \S 9.

\defcitealias{wesson:05}{W05} 

\begin{table*}
\label{tab:PNe}
	\centering
	\caption{Characteristics of analysed PNe}
	\begin{tabular}{llcccccccl} 
		\hline
		Object &  Name & log(F(H$\beta$))$^a$ & Distance$^b$ & Size & Morph.$^c$ & V$_{rad}$& ADF$^a$(O$^{+2}$)& Star$^d$ & Comments\\
		       &       & erg cm$^{-2}$s$^{-1}$&  (kpc)  & arcsec &             & (km s$^{-1}$)&  \citetalias{wesson:05} & & References\\
		\hline \hline 
		PN G038.2+12.0     & Cn\,3-1  & $-$10.94  & 4.68    & 4.8       &  Ec t * h & $-$11.8   & ---   & O(H)7Ib       & IRAS18152+1007\\
		  \hspace{1cm}   ''            &          &           & 7.68    &           &           &           &       &               & Gaia DR3 \\
		PN G045.4-02.7$^e$ & Vy\,2-2  & $-$11.56  & 3.51    & 3.1x2.6   & P         & $-$71.2   & 11.8  & B[e]          &  \citet{lamers:98}\\
		PN G051.4+09.6     & Hu\,2-1  & $-$10.80  & 4.18    & 2.6       & Bc bcr(o) & +17.0     & 4.0   &  WNb+?        & \\
		  \hspace{1cm}   ''             &          &           & 0.27   &           &           &           &       &               & Hipparcos distance\\
		PN G053.3+24.0$^f$     & Vy\,1-2  & $-$11.53  & 8.13    & 4.6       &B fastOut  &           & 6.17  & [WR]/$wels$ &  \\
		PN G058.3-10.9     & IC\,4997 & ---       & 4.24    & 7.2       & Bc t?     & +07.2     & ---   & $wels$        & \\ 
		\hline
		\multicolumn{9}{l}{$^a$ Spectroscopy analysed by \citet{wesson:05} \citepalias{wesson:05}.} \\
		\multicolumn{9}{l}{$^b$ Distances from \citet{frew:16}, unless otherwise specified.}\\
		\multicolumn{9}{l}{$^c$ Morphology mostly from \citet{sahai:11}.}\\
		\multicolumn{9}{l}{$^d$ Star classification from \citet{weidmann:20}. For Vy\,1-2 it is from \citet{lamers:98}.}\\
		\multicolumn{9}{l}{$^e$ Wide H$\alpha$ line due to Raman Effect, reported by \citet{arrieta:03}.}\\
		\multicolumn{9}{l}{$^f$ A deep study of this nebula can be found in \citet{akras:15}}\\
	\end{tabular}
\end{table*}

\section{Observations and data reduction}

To derive the chemical abundances from ORLs, high resolution spectra ($\delta\lambda$  better than 0.5 \AA/pix) are required because these lines appear generally blended with other ORLs or stellar lines.
Therefore, for data acquisition we used the highest spectral resolution instrument available.

Some of the objects presented here have been observed by us since 2001, with the echelle REOSC spectrograph, attached to the 2.1-m telescope of the Observatorio Astron\'omico Nacional, at San Pedro M\'artir, B.C., M\'exico, OAN-SPM. Most recent observations were carried out during September 2018, June and August 2019, with the same instrument, covering the wavelength range from about 3,600 \AA ~to 7,000 \AA. In all cases, the slit was located through the centre  of the nebula crossing the central star position, and oriented E-W (P.A. 90$^\circ$).  The slit length was always 13.3 arcsec in the plane of the sky. The slit widths used were 100 $\mu$ (1.3 arcsec)  and 150 $\mu$ (2 arcsec), which provided spectral resolutions $\rm{R} = \lambda / \delta \lambda$ of 20,000 and 18,000, respectively, at 5,000 \AA. Long and short exposure times were used in order to detect the weak lines (in particular the heavy-element recombination lines) with good signal to noise and  to not saturate the intense lines (exposure times of 1,800 s, 900 s, 600 s or less were employed).

Wavelength calibration was performed with a Th-Ar lamp obtained after each scientific observation. Standard stars from the list by \citet{hamuy:92} were observed each night for flux calibration. The slit width for standard stars was of 300 $\mu$ (about 4 arcsec), in order to include all the stellar flux in the slit.

The log of observations is presented in Table \ref{table:observations}. Observations obtained since 2001 are included. 
\begin{table*}
	\centering
	\caption{Log of observations}
	\begin{tabular}{lcccccccc} 
		\hline 
		Object  &  Name & Obs. Date & Exp. Time & Slit Size\\
		        &       &           & (s)       & arcsec\\
		\hline \hline 
		PN G038.2+12.0 & Cn\,3-1  & 15/08/2019 & 3$\times$1800, 900 & 2\\
		PN G045.4-02.7 & Vy\,2-2  & 30/06/2019 & 3$\times$1800      & 2\\
		PN G051.4+09.6 & Hu\,2-1  & 13/08/2019 & 2$\times$1800, 600 & 2\\
		PN G053.3+24.0 & Vy\,1-2  & 21/09/2018 & 2$\times$1800      & 1.3 \\
		PN G058.3-10.9 & IC\,4997 & 25/08/2001 & 900, 60            & 4\\ 
		\hline
	\end{tabular}
	\label{table:observations}
\end{table*}

\subsection{Data reduction}
Spectroscopic data were reduced following the \textsc{iraf}\footnote{IRAF is distributed by the National Optical Astronomy Observatories, which is operated the Association of Universities for Research in Astronomy, Inc., under contract to the National Science Foundation.} standard routines: set of BIAS obtained each night was combined in a MASTER-BIAS for BIAS subtraction. No flat field calibration was carried out. Spectra of the same object, obtained during a night, with the same exposure time, were combined in one, to improve the signal to noise. 1D spectra were extracted from the 2D echelle orders, they were wavelength calibrated with the Th-Ar lamp obtained after each observation and corrected by effects of atmospheric extinction, using the curve for OAN-SPM sky given by \citet{schuster:01}. Finally the flux calibration was carried out by using the standard stars spectra. 

\begin{figure*}
	\includegraphics[width=\columnwidth]{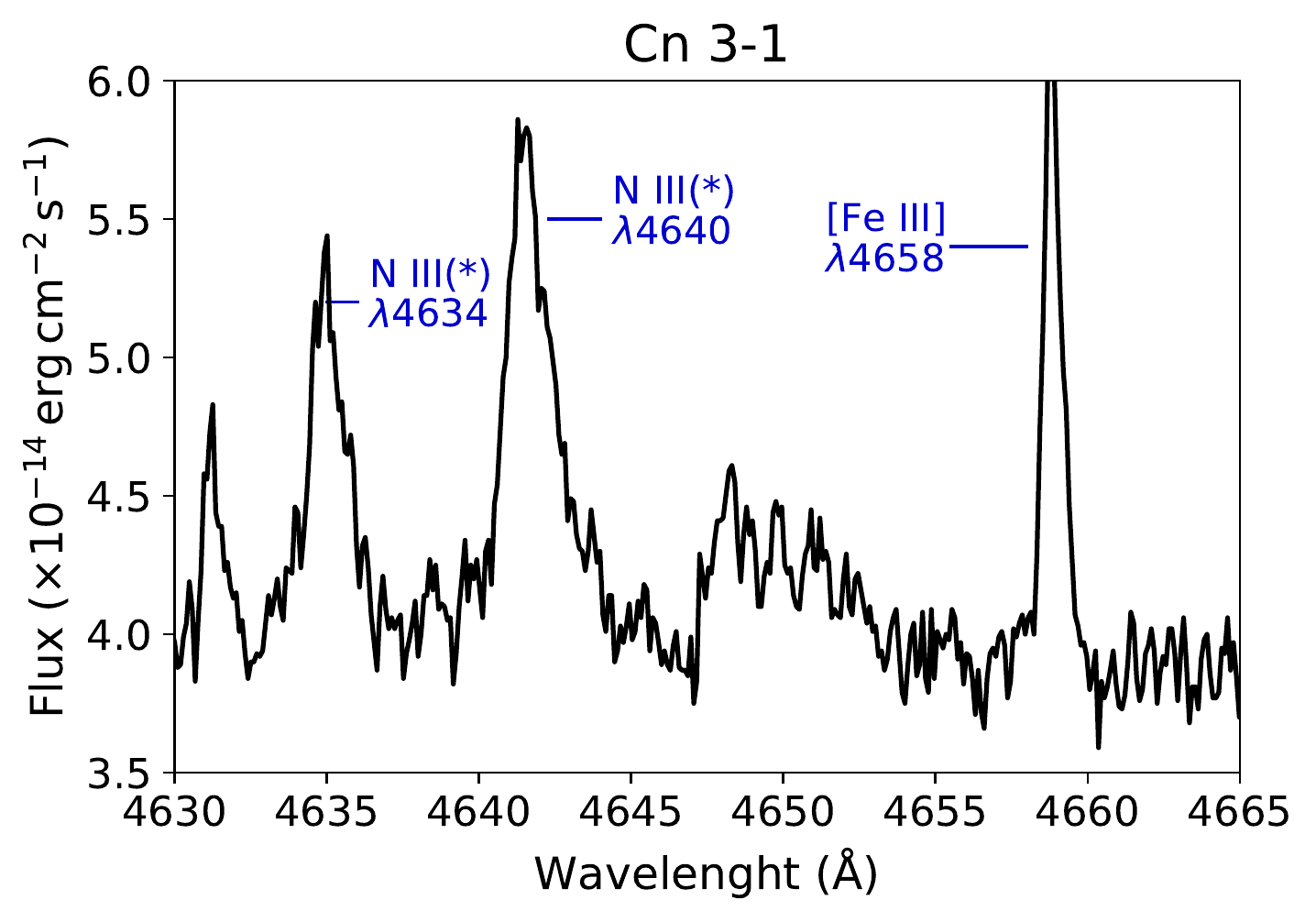}
	\includegraphics[width=\columnwidth]{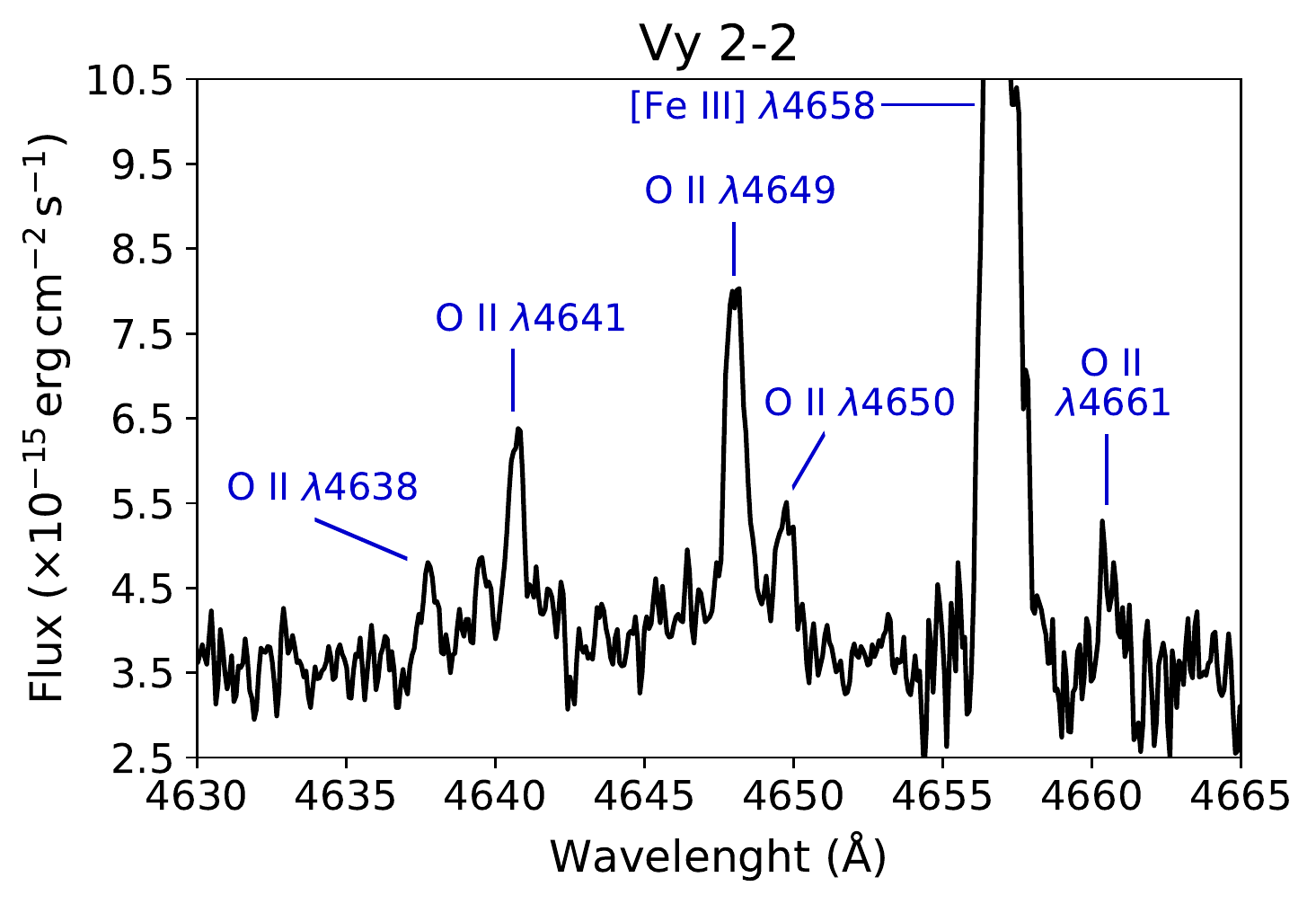}
	\includegraphics[width=\columnwidth]{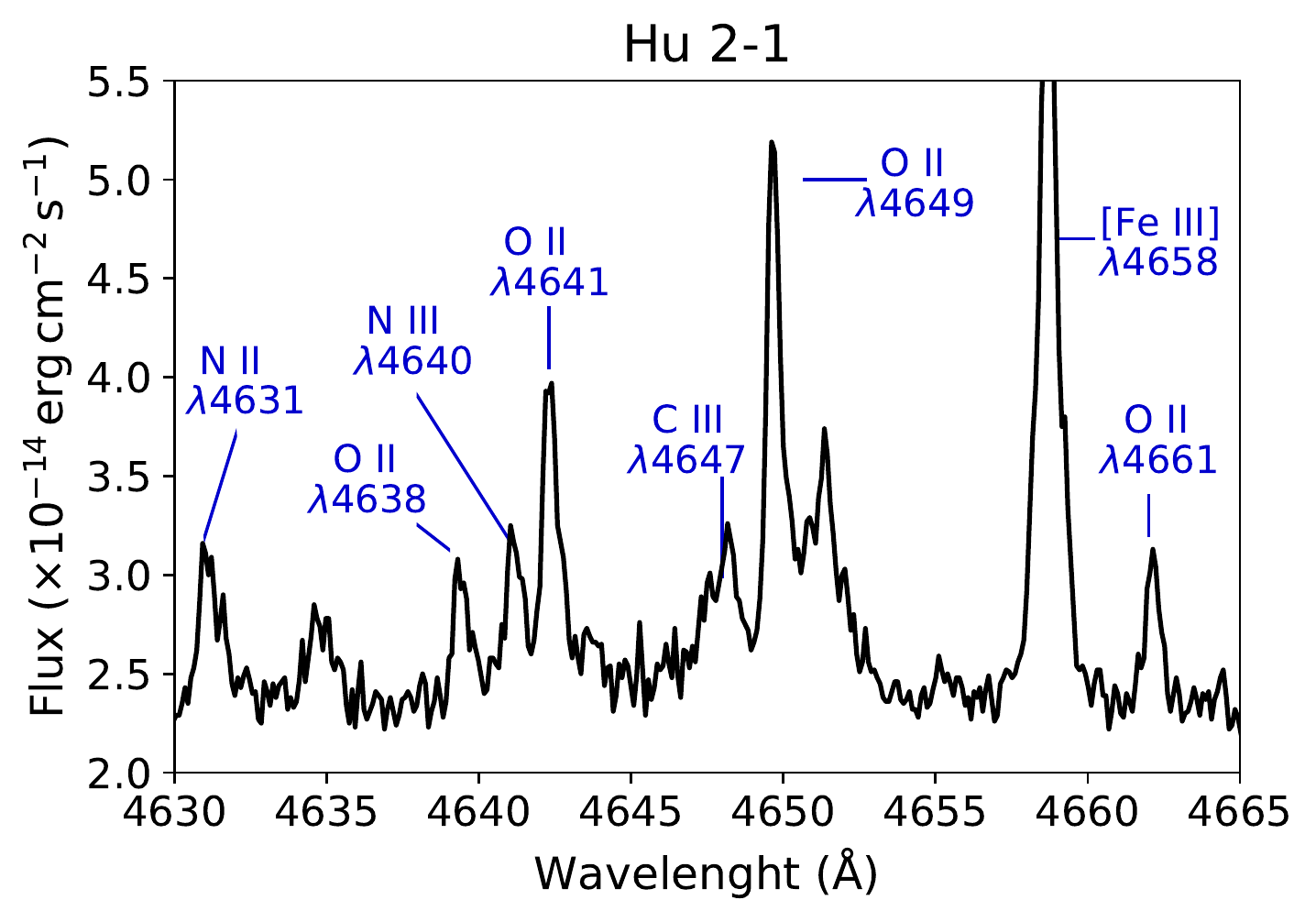}
	\includegraphics[width=\columnwidth]{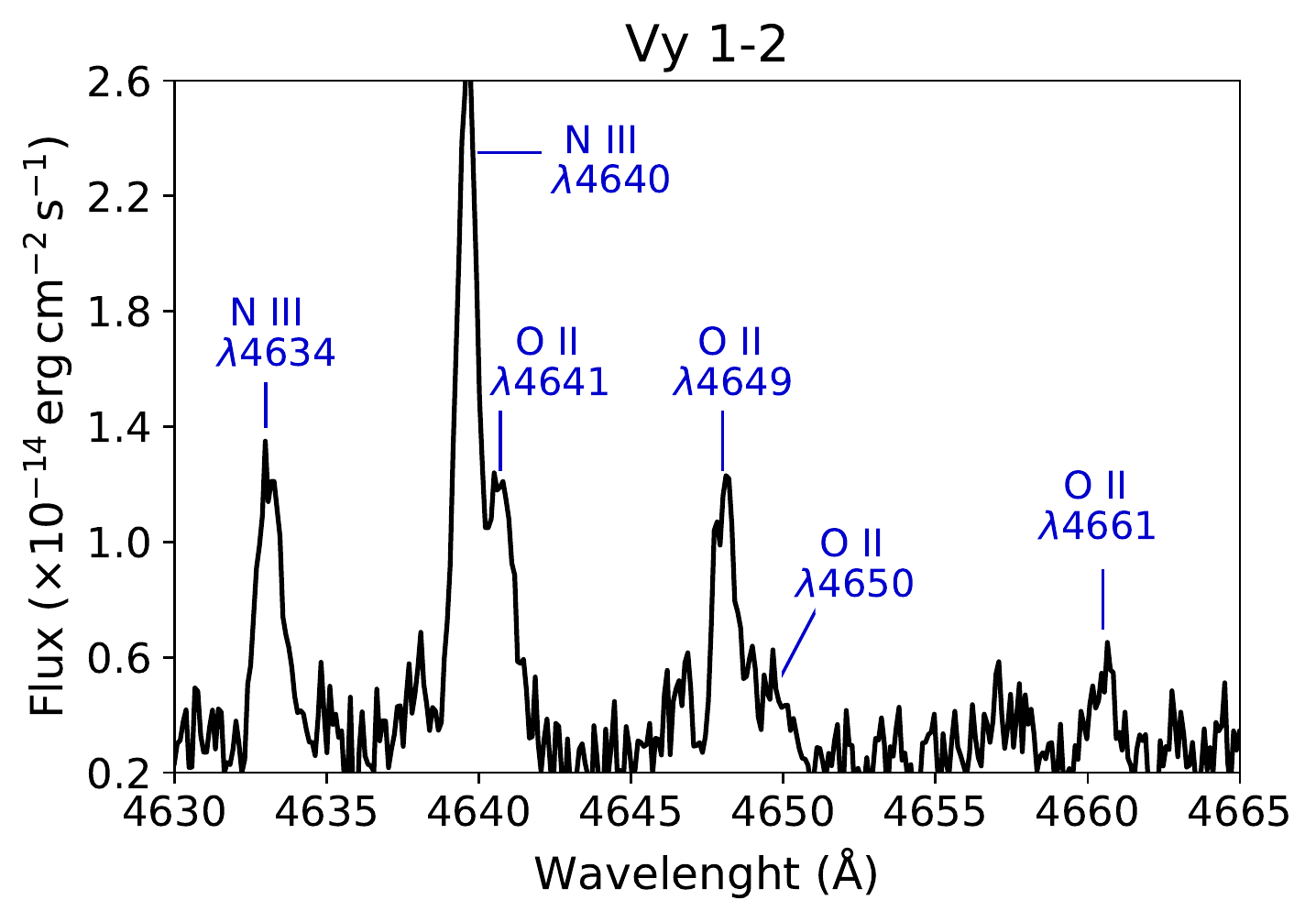}
	\includegraphics[width=\columnwidth]{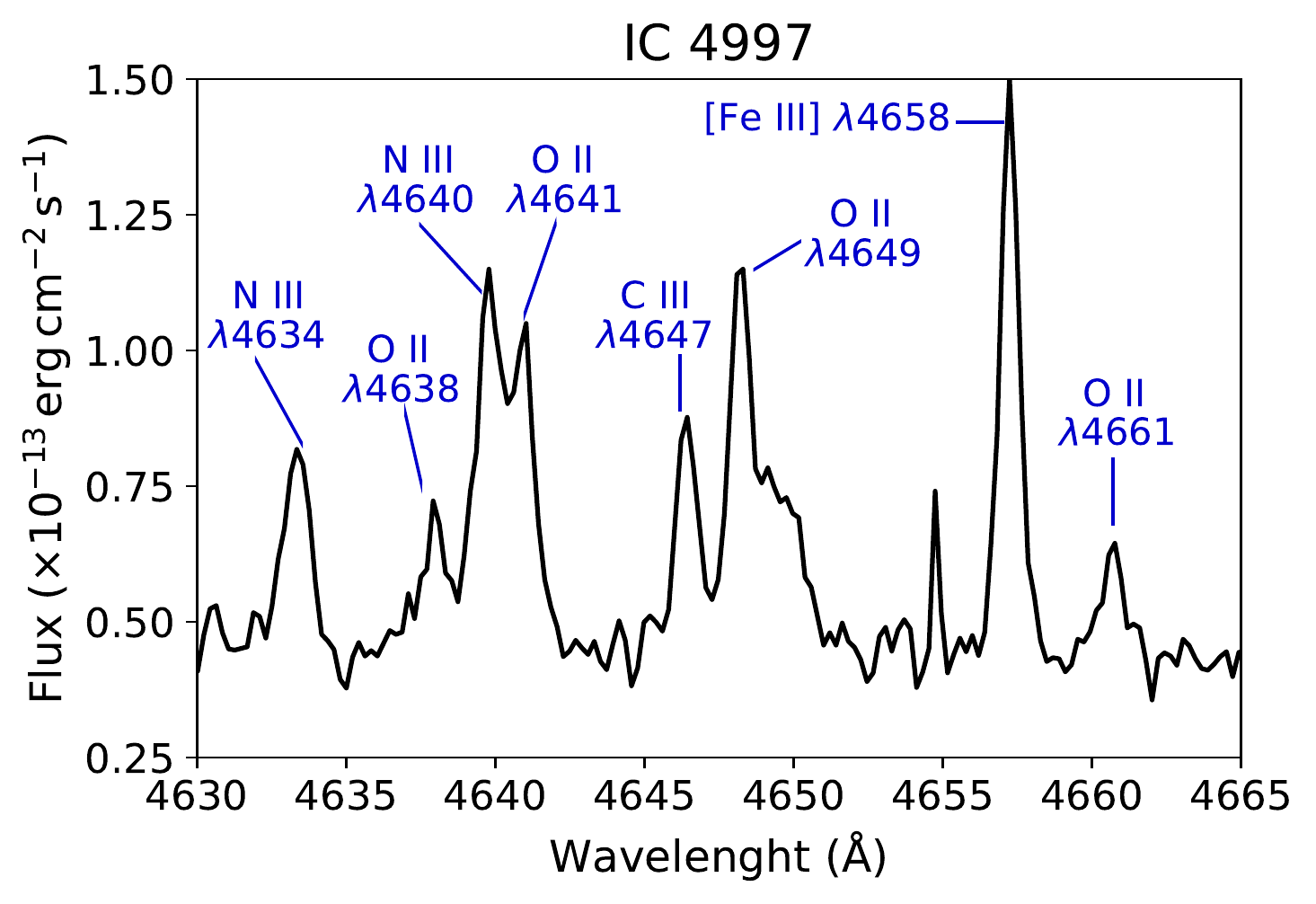}
    \caption{Section of the calibrated spectra of the sample, around 4650 \AA. Important recombination lines of heavy elements are highlighted. An asterisk in the line identification indicates that the line is emitted by the central star and not by the nebula.} 
    \label{fig:spectra}
\end{figure*}

Spectra in the zone around $\lambda$4650 are presented in Fig. \ref{fig:spectra} showing the quality of our data and the adequate spectral resolution to analyse the lines in this zone, which is very important to separate the \ion{O}{ii} and \ion{N}{ii} ORLs. 

Fluxes were measured for all the available lines (which amount to more than 100 in each case), by using \textsc{iraf}'s \textit{splot} task. Their "Full Width at Half Maximum", FWHM, were determined by applying a Gaussian fit in each case. If a line was double (two peaks) each component was measured and a Gaussian fit was applied to each one. Such lines are marked with an asterisk in the intensity tables. 
FWHMs were corrected by effects of instrumental width (derived from 
the line widths of comparison lamps) and thermal broadening (by adopting an electron temperature according to each case), by assuming they add in quadrature.  Turbulent velocities were not considered (see \S7).

  Line flux errors were calculated by using the \citet{tresse:99}
expression shown here:
\begin{equation}
\rm{ \sigma_{F} = \sigma_{c}D\sqrt{2N_{pix}+\frac{EW}{D}} }
\end{equation}
 where D is the spectral dispersion in \AA/pix, $\rm{\sigma_c} $ is the mean standard deviation per pixel of the continuum measured on each side
of the line, and $\rm{ N_{pix} }$ is the number of pixels covered by the line. For the upcoming calculations, we adopted $\rm{2\sigma_{F}}$ values as more adequate to represent the true line errors.

For each observed object, the logarithmic reddening correction, c(H$\beta$), was derived by using \citet{cardelli:89} reddening law, assuming a ratio of total to selective extinction $\rm{R_V} = 3.1$ and by using the different theoretical ratios $\rm{H\alpha / H\beta, \, H\gamma / H\beta, \, H\delta / H\beta \, and \, H\epsilon / H\beta}$ given by \citet{storey:95} for the temperatures and densities presented for each object in Table \ref{tab:dereddening}. H$\alpha$ line was saturated in the spectra of Hu\,2-1 and IC\,4997, even in their low exposure-times spectra, therefore the $\rm{H\alpha / H\beta}$ ratio was not used.

\begin{table} 
\centering
\caption{Logarithmic reddening correction and assumed temperatures and densities for the theoretical Balmer ratios used in the calculation}
\begin{tabular}{lcccc}\hline
Object & c(H$\beta$) & $\rm{T_e}$ & $\rm{log(n_e)}$  & $\rm{H / H\beta}$ (avg) \\
       &           & ($10^4$ K)  &  (cm$^{-3}$) & \\
       \hline \hline 
Cn\,3-1 & 0.65$\pm$0.07 & 0.75 & 4 & $\rm{H\alpha, H\gamma, H\delta, H\epsilon}$\\
Vy\,2-2 & 0.66$\pm$0.01 & 1.00 & 4 & $\rm{H\alpha}$ \\
Hu\,2-1 & 0.45$\pm$0.05 & 1.00 & 4 & $\rm{H\gamma, H\delta}$ \\
Vy\,1-2 & 0.31$\pm$0.06 & 1.00 & 4 & $\rm{H\alpha, H\gamma, H\delta}$ \\
IC\,4997 & 0.32$\pm$0.01 & 1.50 & 6 & $\rm{H\gamma}$ \\
\hline
\end{tabular}
\label{tab:dereddening}
\end{table}

The observed and de-reddened fluxes and the corrected line widths for each object are presented in individual tables. Such tables can be found in the on-line version only. An example of such tables is presented in Table \ref{table:fluxes} for Cn\,3-1, where the emitting ion, the wavelength at rest, the observed central wavelength, the observed and de-reddened fluxes, relative to H$\beta$, and their uncertainties obtained by propagating the errors associated to the line fluxes through the normalization to H$\beta$ and the de-reddening correction, are presented together with the line widths (FWHM) and the expansion velocities derived from the line widths.  The individual tables of all PNe are available as online supplementary data. 

\begin{table*}
\centering
\caption{Example: Cn\,3-1: observed (F$_{\lambda}$)  and de-reddened (I$_{\lambda}$) fluxes normalized to $\rm{H\beta = 100}$, their absolute errors, line widths FWHM and expansion velocities. Tables for all objects are available as on-line material}
\label{table:fluxes}
\begin{tabular}{cccccccccc} \hline
Ion & $\lambda_0$ & $\rm{\lambda_{obs}}$ & $\rm{F_{\lambda}/F(H\beta)}$ & $\rm{\delta(F_{\lambda})}$ & $\rm{I_{\lambda}/I(H\beta)}$ & $\rm{\delta(I_{\lambda})}$ & $\rm{V_{rad} \, hel}$ & FWHM & $\rm{v_{exp}}$ \\ 
 & (\AA) & (\AA) & & & & & $(\rm{km \, s^{-1}})$ & (\AA) & $\rm{(km \, s^{-1})}$ \\ \hline \hline
\ion{He}{i} & 3634.25 & 3634.84 & 0.16 & 0.02 & 0.26 & 0.03 & 30.14 & 0.32 & 13.39 \\
H29 & 3663.40 & 3663.98 & 0.15 & 0.02 & 0.24 & 0.04 & 29.18 & 0.54 & 22.18 \\
H28 & 3664.68 & 3665.21 & 0.19 & 0.02 & 0.31 & 0.03 & 24.99 & 0.39 & 15.81 \\
H27 & 3666.10 & 3666.65 & 0.23 & 0.02 & 0.38 & 0.03 & 26.78 & 0.31 & 12.50 \\
H26 & 3667.68 & 3668.21 & 0.22 & 0.02 & 0.36 & 0.04 & 24.96 & 0.40 & 16.37 \\
H25 & 3669.46 & 3669.97 & 0.28 & 0.02 & 0.46 & 0.04 & 23.71 & 0.48 & 19.77 \\
H24 & 3671.48 & 3672.04 & 0.29 & 0.02 & 0.48 & 0.03 & 27.36 & 0.59 & 24.13 \\
H23 & 3673.76 & 3674.28 & 0.43 & 0.01 & 0.69 & 0.04 & 24.31 & 0.49 & 19.94 \\
H22 & 3676.36 & 3676.93 & 0.40 & 0.02 & 0.65 & 0.04 & 27.88 & 0.43 & 17.54 \\
H21 & 3679.36 & 3679.88 & 0.40 & 0.02 & 0.66 & 0.04 & 24.17 & 0.47 & 19.33 \\
H20 & 3682.81 & 3683.38 & 0.62 & 0.02 & 1.00 & 0.05 & 28.04 & 0.52 & 21.18 \\
H19 & 3686.83 & 3687.42 & 0.58 & 0.03 & 0.94 & 0.06 & 29.53 & 0.50 & 20.38 \\
H18 & 3691.56 & 3692.15 & 0.74 & 0.03 & 1.19 & 0.07 & 29.72 & 0.55 & 22.32 \\
H17 & 3697.15 & 3697.76 & 0.85 & 0.02 & 1.38 & 0.07 & 31.27 & 0.54 & 21.90 \\
H16 & 3703.86 & 3704.42 & 0.81 & 0.02 & 1.32 & 0.07 & 27.37 & 0.46 & 18.46 \\
\ion{He}{i} & 3705.04 & 3705.57 & 0.16 & 0.02 & 0.25 & 0.04 & 24.28 & 0.47 & 18.88 \\
H15 & 3711.97 & 3712.53 & 1.18 & 0.02 & 1.91 & 0.09 & 27.19 & 0.46 & 18.57 \\
H14 & 3721.83 & 3722.45 & 1.66 & 0.03 & 2.69 & 0.13 & 31.83 & 0.50 & 20.23 \\
$[\ion{O}{ii}]$ & 3726.03 & 3726.58 & 66.26 & 0.12 & 107.29 & 4.78 & 26.38 & 0.41 & 16.62 \\
$[\ion{O}{ii}]$ & 3728.82 & 3729.33 & 30.44 & 0.08 & 49.28 & 2.20 & 23.04 & 0.39 & 15.81 \\ 
... \\
		\hline
		\end{tabular}
		\end{table*}
		
\section{Physical parameters of nebulae}

From the nebular lines, in particular from those collisionally excited lines, physical conditions, such as electron densities and temperatures, can be derived from some diagnostic line ratios.  Densities can be determined from the [\ion{S}{ii}] $\lambda\lambda$6731/6716, [\ion{O}{ii}] $\lambda\lambda$3729/3726, [\ion{Cl}{iii}] $\lambda\lambda$5538/5518,  [\ion{Fe}{iii}] $\lambda\lambda$4701/4659, and [\ion{Ar}{iv}] $\lambda\lambda$4711/4740 intensity ratios. Electron temperatures can be obtained from the [\ion{N}{ii}] $\lambda\lambda$(6548+6584)/5755, [\ion{O}{iii}] $\lambda\lambda$(5007+4959)/4363, [\ion{Ar}{iv}] $\lambda\lambda$(7170+7263)/(4711+4740),  [\ion{S}{ii}] $\lambda\lambda$(6716+6731)/(4068+4076) and [\ion{O}{ii}] $\lambda\lambda$7325/3727 intensity  ratios.

From the available diagnostic line ratios, all densities and temperatures were calculated with the code \textsc{pyneb} \citep{luridiana:15}, using the atomic data presented in the Appendix A, Table \ref{tab:atomic-parameters}. \textsc{pyneb} routine \textit{getCrossTemden} was used to determine simultaneously the temperature and density from the [\ion{N}{ii}], [\ion{O}{iii}], [\ion{S}{ii}] and other diagnostic lines, by building diagnostic diagrams. The uncertainties in the physical conditions and abundances were estimated using Monte Carlo simulations, with 400 random points following a normal distribution around the observed intensity of the lines.

\subsection{ Diagnostics diagrams}
\begin{figure*}
    \centering
	\includegraphics[width=\columnwidth]{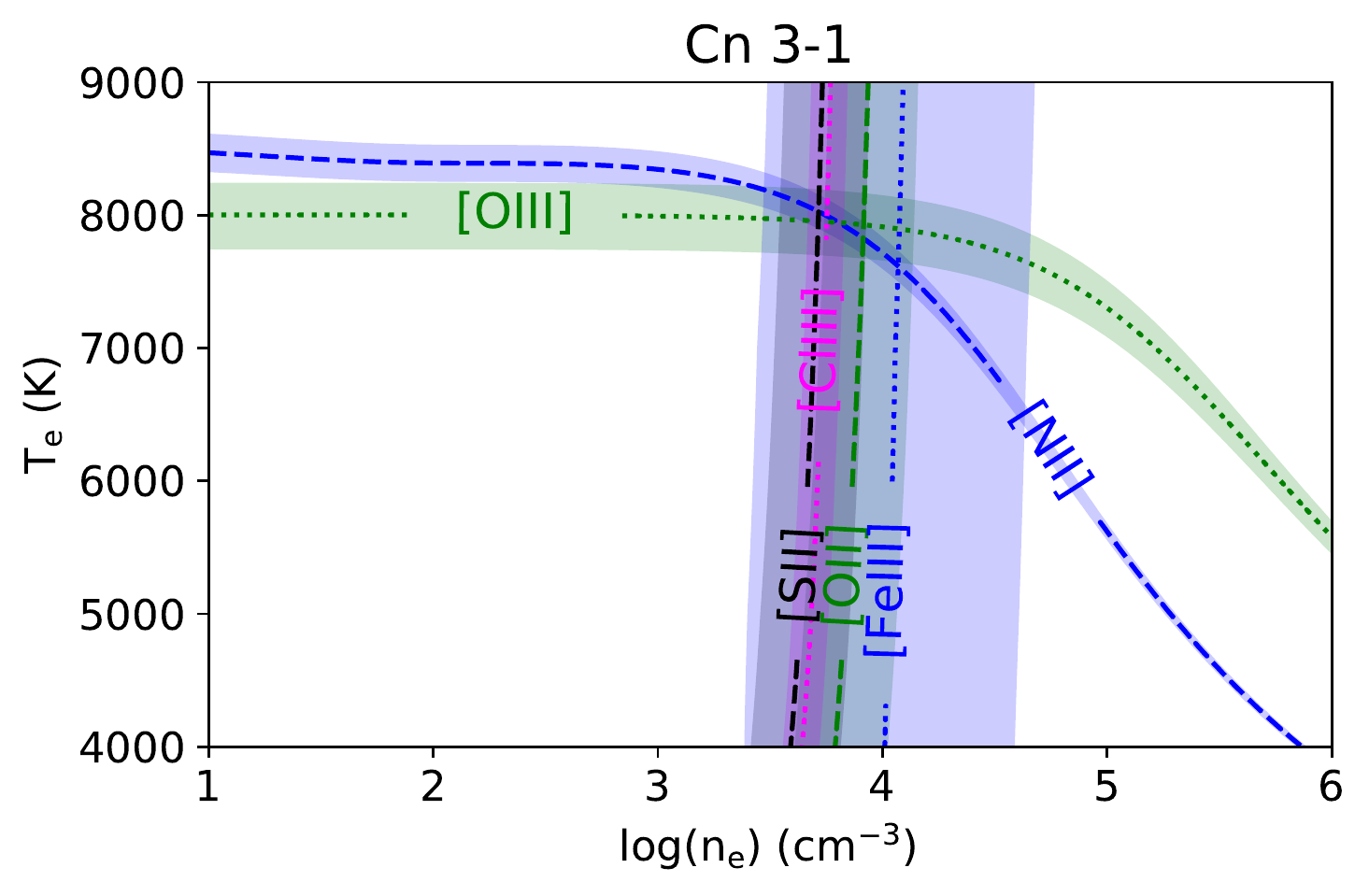}
	\includegraphics[width=\columnwidth]{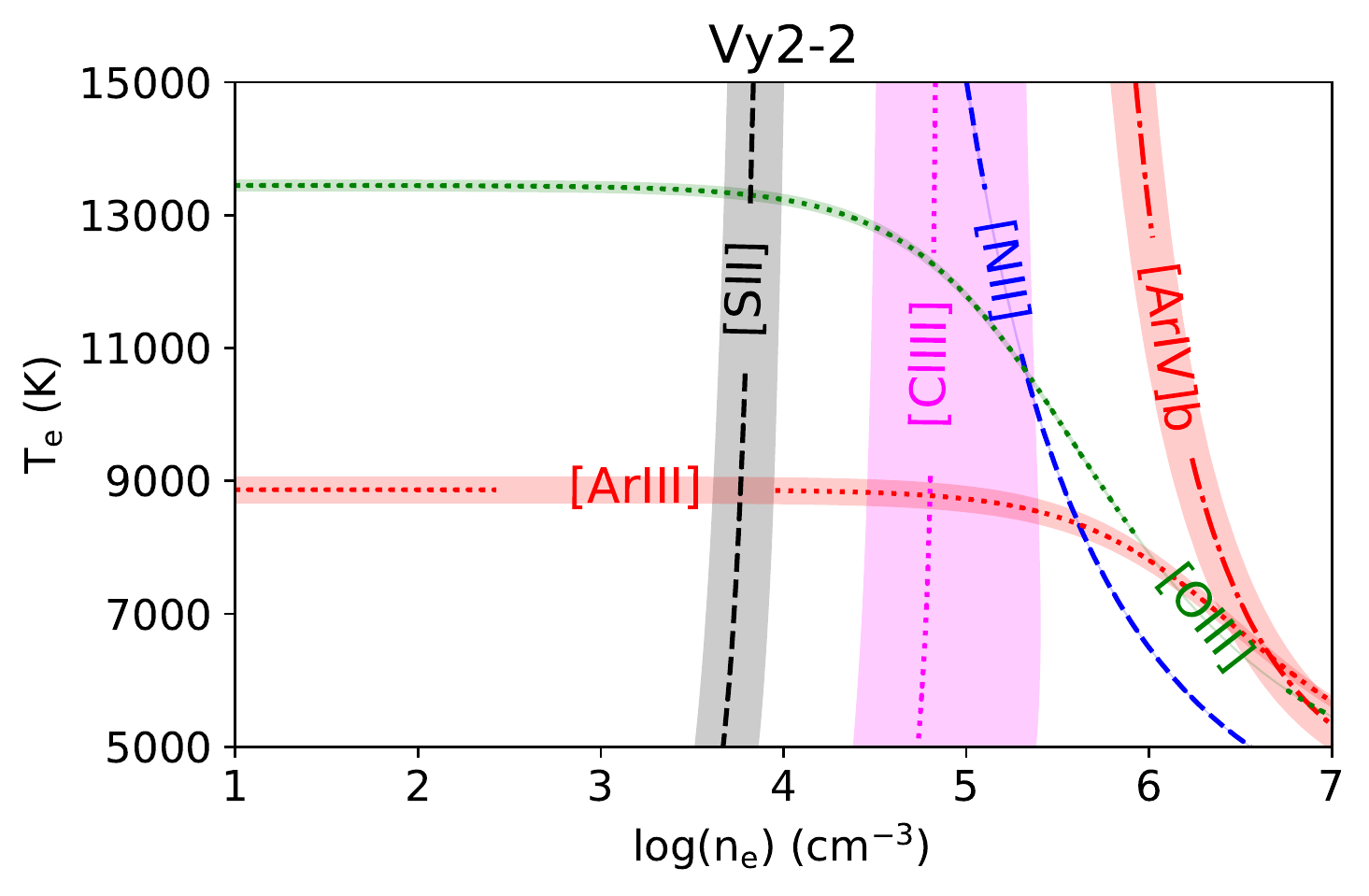}
	\includegraphics[width=\columnwidth]{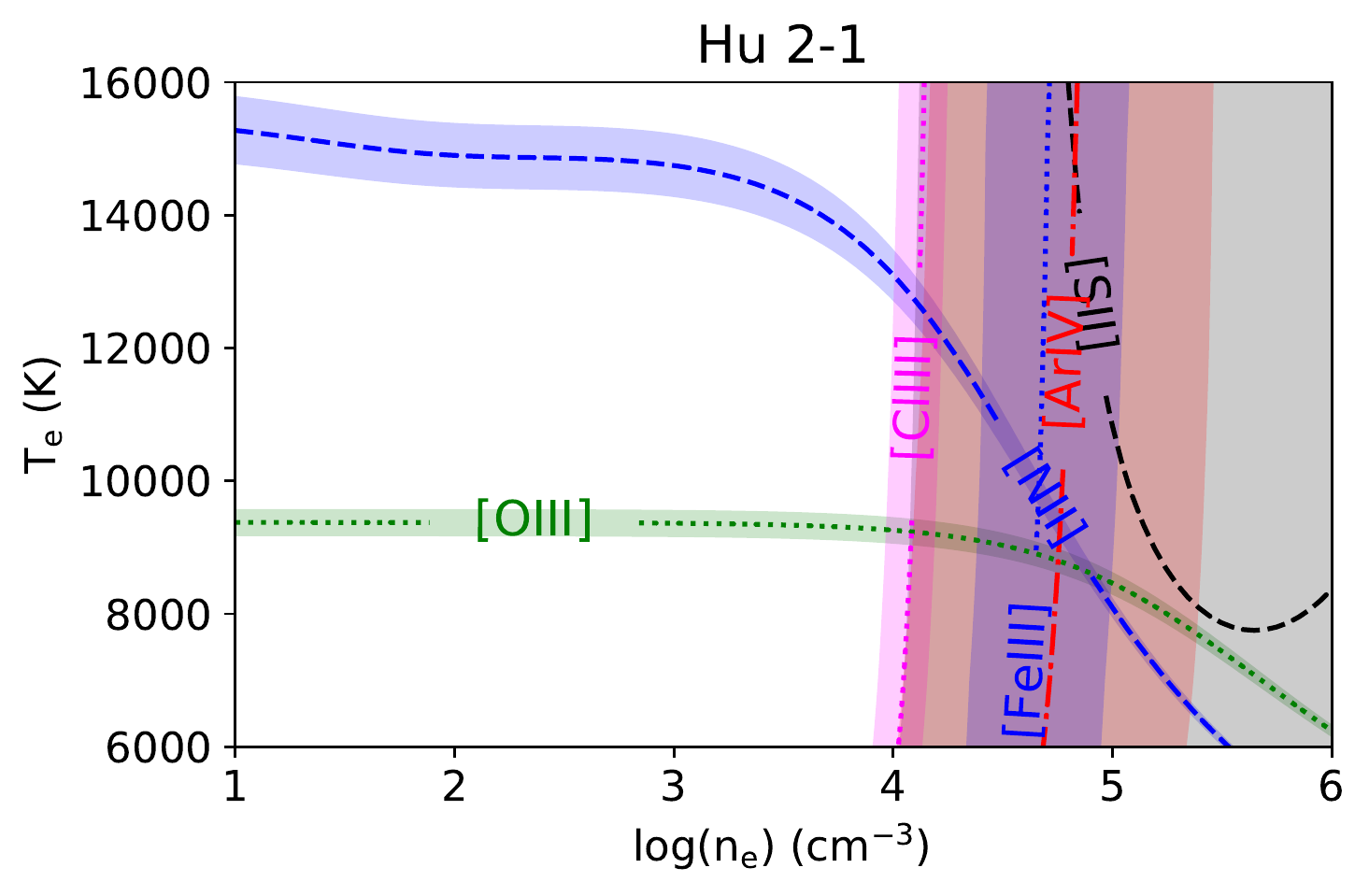}
	\includegraphics[width=\columnwidth]{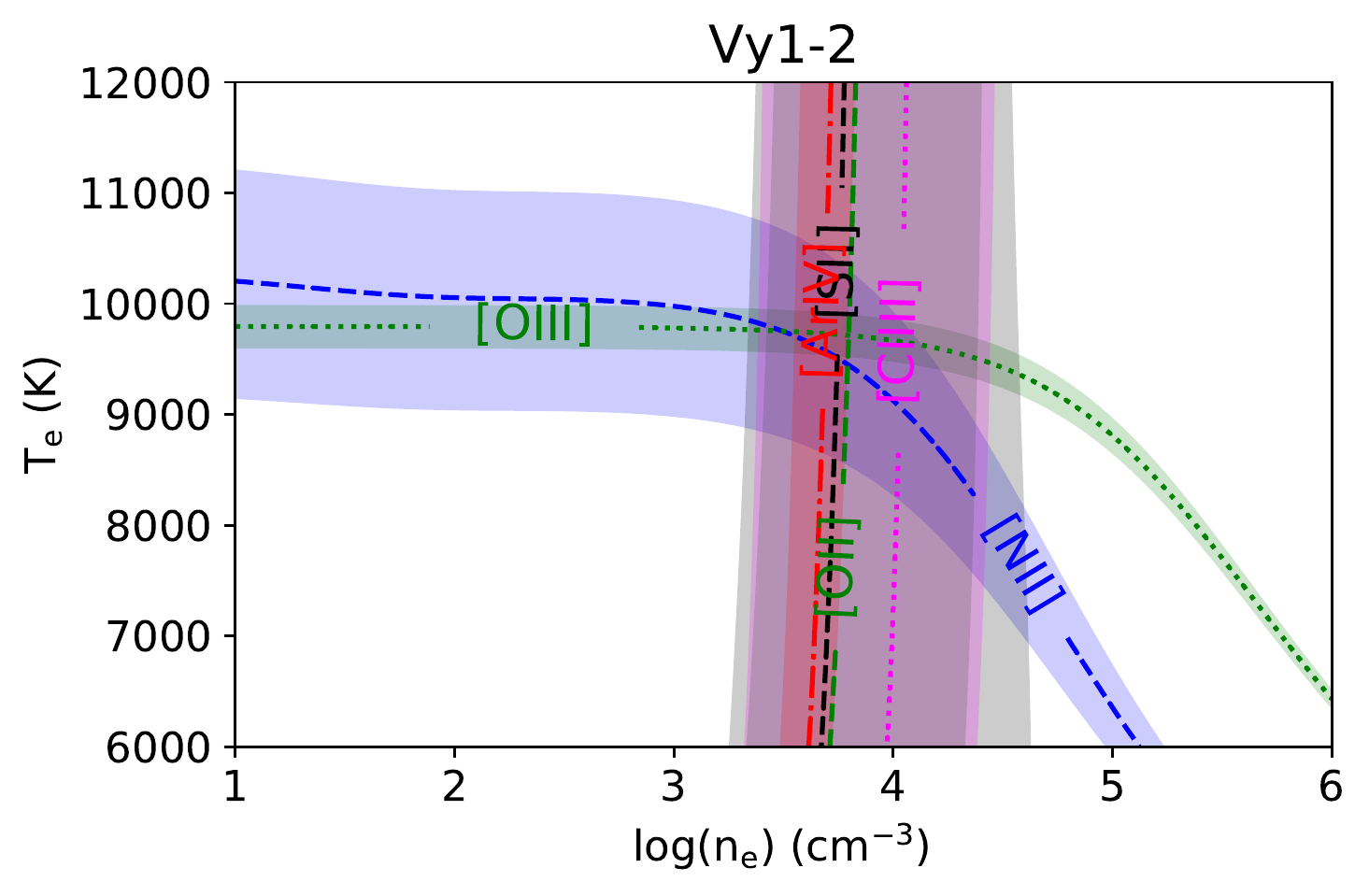}
	\includegraphics[width=\columnwidth]{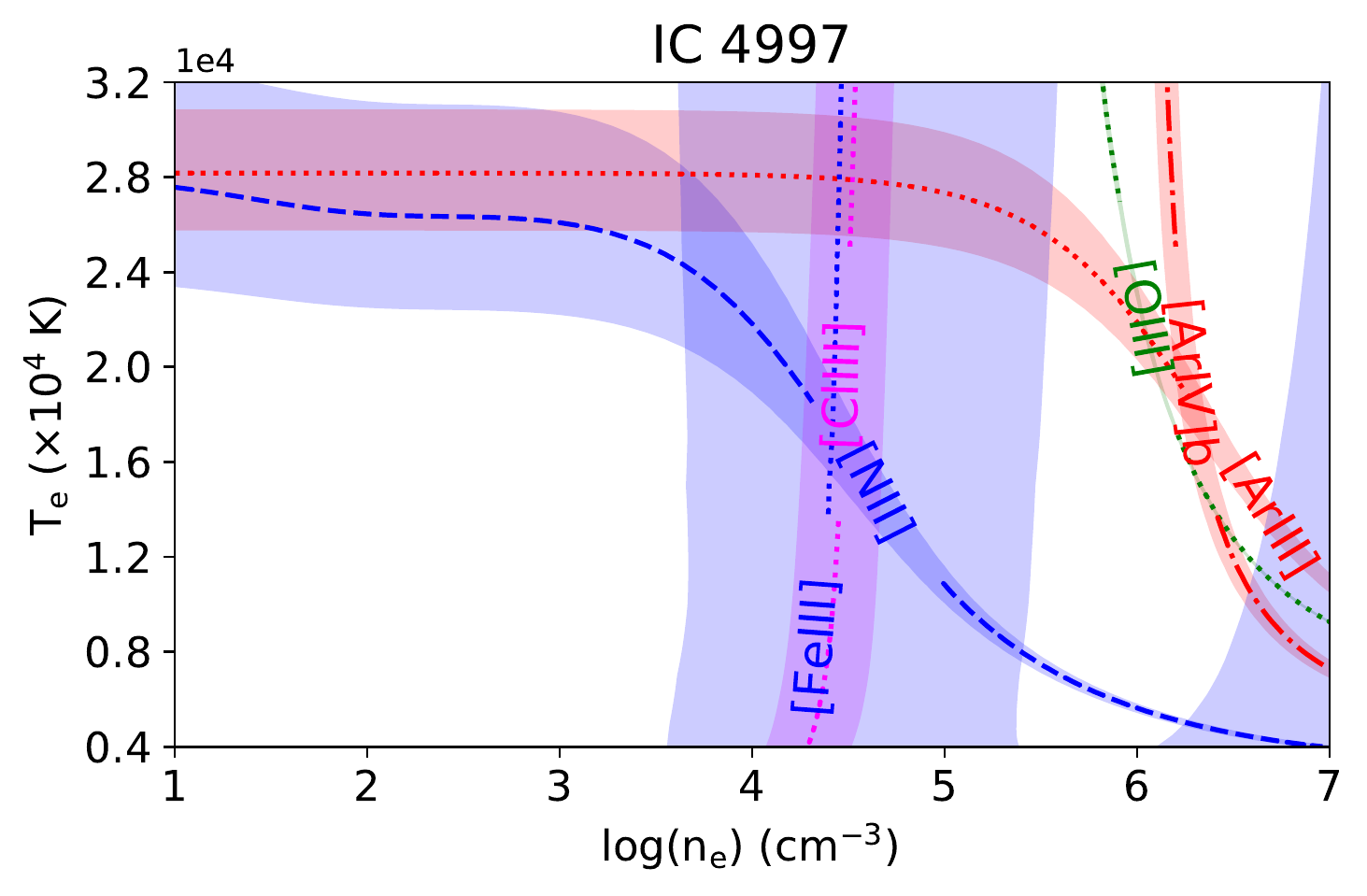}
    \caption{Diagnostic diagrams for density and temperature, derived with \textsc{pyneb}, for the different objects. Sensitive CELs line ratios are indicated. 
    Coloured shadowed bands represent the 1 sigma rms error of each diagnostic. [\ion{Ar}{iv}]b represents the line ratio $\lambda \lambda$7170/4740.} 
    \label{fig:diagnostics}
\end{figure*}

In Fig. \ref{fig:diagnostics} diagnostic diagrams for the analysed objects are presented.  These diagrams show the behaviour of diagnostic line ratios with the electron density and temperature. Each line ratio is represented by a broken or dotted line inside a colour band  which shows the 1$\sigma$ rms error. Usually the electron densities and temperatures are obtained from the zone where density- and temperature-diagnostics intersect. The  derived values for our objects are listed in Table \ref{table:density-temperature}.
These  temperatures and densities will be used to determine ionic abundances.

 In order to derive correct temperatures, intensities of the auroral lines [\ion{N}{ii}] $\lambda$5755 and [\ion{O}{iii}] $\lambda$4363 were corrected by effects of recombination using \citet{liu:00} relations:
\begin{equation}
\rm{ \frac{I_R(\lambda 5755)}{I(H\beta)} = 3.19\,t^{0.30} \times \frac{N^{+2}}{H^+} },
\end{equation}
\noindent where $\rm{t = T_e([\ion{N}{ii}])/10^4 K}$, 
and $\rm{\frac{N^{+2}}{H^+} }$ is the abundance determined from recombination lines.

For [\ion{O}{iii}]$\lambda$4363, the expression used is:
\begin{equation}
\rm{ \frac{I_R(\lambda 4363)}{I(H\beta)} = 12.4 \,t^{0.59} \times \frac{O^{+3}}{H^+}, }
\end{equation}
where $\rm{t = T_e([\ion{O}{iii}])/10^4 \, K}$ and $\rm{ \frac{O^{+3}}{H^+}}$ is the abundance determined from recombination lines. This correction was only applied to Vy\,1-2 which is the only highly ionized object of the sample. As the observed lines of \ion{O}{iii} are emitted by Bowen resonance and not by pure recombination \citep{grandi:76, garcia-rojas:13}, we could estimate $\rm{ \frac{O^{+3}}{H^+} }$ using the next relation, given by \citet{kingsburgh:94}:
\begin{equation}
\rm{ \frac{ O^{+3}}{H^+}  = \left[ \left( \frac{He}{He^+} \right) - 1 \right] \times \left( \frac{O^{+}}{H^+} + \frac{O^{+2}}{H^+}.  \right) }
\end{equation}

In this case we used the O$^+$ and O$^{+2}$ abundances determined from CELs because we were not able to determine the abundance of O$^+$ from recombination lines. 

For Vy\,1-2, we determined that the recombination contribution to auroral line [\ion{O}{iii}] $\lambda4363$ intensity is of $\sim$1\%. We also found the same percentage using the recent relation proposed by \citet{gomez-llanos:20a} for this correction. The $\mathrm{T_e}$ determined using the uncorrected auroral line intensity is about $\sim$200 K higher than that determined using the corrected line intensity, which lies within the error bars. Therefore the correction does not produce a significant effect in abundances determinations. In addition, \citet{garcia-rojas:12} were able to determined O$^{+3}$ abundance directly from ORLs for the highly ionized PNe of their sample, they found that the recombination contribution is about $\sim$3\% and the correction was negligible in the temperature determinations.

\begin{table*}
\centering
    \caption{Electron densities and temperatures$^{(a)}$ from collisionally excited lines}
    \begin{tabular}{lcccccc} \hline
      & Line ratio & Cn\,3-1 & Vy\,2-2 & Hu\,2-1 & Vy\,1-2 & IC\,4997 \\ \hline \hline 
    $\rm{n_e}$ [\ion{S}{ii}] & 6731/6716 & 5,100$_{-1,500}^{+3,000}$ & 6,200$_{-1,800}^{+3,000}$ & --- & 4,600$_{-2,300}^{+8,200}$ & --- \\
    $\rm{n_e}$ [\ion{O}{ii}] & 3729/3726 & 7,900$_{-2,400}^{+5,000}$ & --- & --- & 5,100$_{-2,500}^{+8,700}$ & --- \\
    $\rm{n_e}$ [\ion{Cl}{iii}] & 5538/5518 & 5,600$_{-800}^{+1,100}$ & 58,200$_{-23,200}^{+76,600}$ & 13,200$_{-2,700}^{+4,200}$ & 9,300$_{-5,300}^{+15,800}$ & 31,700$_{-11,200}^{+21,100}$ \\
    $\rm{n_e}$ [\ion{Fe}{iii}] & 4701/4659 & 10,500$_{-6,700}^{+16,200}$ & --- & 32,300$_{-11,200}^{+53,900}$ & --- & 23,200$_{-18,200}^{+340,500}$ \\
    $\rm{n_e}$ [\ion{Ar}{iv}] & 4740/4711 & --- & --- & 48,700$_{-27,100}^{+88,000}$ & 4,800$_{-1,200}^{+1,400}$ & --- \\ \hline
    \textit{Adopted} &  & 6,600$_{-1,400}^{+2,100}$ & 1) 6,200$_{-1,800}^{+3,000}$ & 1) 13,200$_{-2,700}^{+4,200}$ & 5,400$_{-1,600}^{+5,200}$ & 1) 31,700$_{-11,000}^{+21,100}$ \\
     &  &  & 2) 58,200$_{-23,200}^{+76,600}$ & 2) 47,100$_{-21,000}^{+79,000}$ &  & 2) 1.81$_{-0.48}^{+0.90}$ $\times 10^6$ $^{(b)}$ \\
     \hline \hline 
    $\rm{T_e}$ [\ion{N}{ii}] & (6584+6548)/5755 & 7,800$\pm$200 & --- & 10,200$_{-1,800}^{+1,100}$ & 9,300$_{-1,500}^{+1,100}$ & 16,200$_{-2,800}^{+3,100}$ \\
    $\rm{T_e}$ [\ion{O}{iii}] & (5007+4959)/4363 & 8,100$\pm$200 & 12,800$\pm$600 & 9,000$\pm$300$^{(c)}$ & 9,700$\pm$200 & --- \\
    $\rm{T_e}$ [\ion{Ar}{iii}] & 7136/5192 & --- & 8,800$\pm$200 & --- & --- & --- \\ \hline
    \textit{Adopted} &  & 7,900$\pm$200 & 1) 8,800$\pm$200 & 9,300$_{-1,500}^{+1,100}$ & 9,500$\pm$500 & 16,200$_{-2,800}^{+3,100}$ \\
    &  &  & 2) 12,800$\pm$600 &  &  & \\ \hline
	\multicolumn{6}{l}{$(a)$: electron densities in particles cm$^{-3}$, temperatures in K}\\
	\multicolumn{6}{l}{$(b)$:  $\rm{n_e}$ from [\ion{O}{iii}] $\lambda \lambda 4959/4363$}\\
	\multicolumn{6}{l}{$(c)$: $\lambda 5007$ line is saturated, $\rm{T_e}$ was determined from $\lambda \lambda 4959/4363$}\\
    \end{tabular}
    \label{table:density-temperature}
\end{table*}

\begin{table*} 
	\centering
	\caption{Adopted zones and their physical conditions for CELs ionic abundance determinations}
    \begin{tabular}{lcccc} \hline
        PN & Zone & Ions & $\rm{n_e (cm^{-3}) }$  & $\rm{T_e (K) }$ \\ \hline \hline 
        Cn\,3-1 & 1 & All observed & 6,600$_{-1,400}^{+2,140}$ & 7,900$\pm$200 \\ \hline
        Vy\,2-2 & 1 & N$^+$, O$^+$, S$^+$, Fe$^+$ & 6,200$_{-1,800}^{+3,000}$ & 8,800$\pm$200 \\
         & 2 & Fe$^{+2}$, S$^{+2}$, Cl$^{+2}$, Ar$^{+2}$ & 58,200$_{-23,200}^{+76,600}$ & 8,800$\pm$200 \\ 
         & 3 & O$^{+2}$, Ne$^{+2}$, Ar$^{+3}$ & 58,200$_{-23,200}^{+76,600}$ & 12,800$\pm$600 \\ \hline
        Hu\,2-1 & 1 & N$^+$, O$^+$, S$^+$, Fe$^+$, S$^{+2}$, Ar$^{+2}$, O$^{+2}$, Cl$^{+2}$ & 13,200$_{-2,700}^{+4,200}$ & 9,300$_{-1,500}^{+1,100}$ \\
        & 2 & Fe$^{+2}$, Ne$^{+2}$, Ar$^{+3}$ & 47,100$_{-21,000}^{+79,000}$ & 9,300$_{-1,500}^{+1,100}$ \\ \hline
        Vy\,1-2 & 1 & All observed & 5,400$_{-1,600}^{+5,400}$ & 9,500$\pm$500 \\ \hline
        IC\,4997 & 1 & N$^+$, O$^+$, S$^+$, Fe$^{+2}$, Cl$^{+2}$, S$^{+2}$ & 31,700$_{-11,000}^{+21,000}$ & 16,200$_{-2,800}^{+3,100}$ \\
        & 2 & O$^{+2}$, Ar$^{+2}$, Ne$^{+2}$, Ar$^{+3}$ & 1.81$_{-0.48}^{+0.90} \times10^6$ & 16,200$_{-2,800}^{+3,100}$ \\ \hline
    \end{tabular}
    \label{table:cel_zones}
\end{table*}

\subsection{Physical parameters from ORLs}

In this section the techniques used to derive electron densities and temperatures from recombination lines are described.
Temperature can be derived from the \ion{He}{i} lines, from the \ion{O}{II} lines, and from the Balmer Jump as explained in the following.

\subsubsection{Temperatures from the Balmer Jump}

Temperature from the Balmer jump, T$\rm{_e}$(BJ), was only determined for Cn\,3-1, Hu\,2-1 and IC\,4997. We followed the procedure given by  \citet{liu:01} for T$\rm{_e}$(BJ) determination:

\begin{equation}
\rm{T_e(BJ) = 368 \times (1+0.259\,y^+ + 3.409\,y^{++})\left( \frac{BJ}{I(H11)} \right)^{-3/2} \, K,}
\end{equation}
where $\rm{ y^+}$ and $\rm{ y^{++}}$ are the ionic abundances He$^+$/H$^+$ and He$^{+2}$/H$^+$, respectively, $\rm{BJ = I_c(3643)-I_c(3681)}$ is the subtraction of intensities of the continuum at those wavelengths, and $\rm{I(H11)}$ is the intensity of \ion{H}{i} $\lambda$3770.

\subsubsection{Temperatures from \ion{He}{i} lines}

Helium electronic temperatures, $\rm{T_e (\ion{He}{i})}$, can be derived using \citet{zhang:05} methodology and the theoretical emissivities for \ion{He}{i} lines by \citet{benjamin:99}. We were able to determine temperatures from the line ratios $\lambda\lambda7281/6678$, $\lambda\lambda7281/5875$, $\lambda\lambda6678/5875$, $\lambda\lambda6678/4471$ and $\lambda\lambda5875/4471$, by assuming a density of 10$^4$ cm$^{-3}$, which agrees with the values derived from recombination lines (see Table \ref{table:recdensity-temperature}). 

 \citet{zhang:05} and \citet{otsuka:10} suggested that the most accurate line ratio to determine $\rm{T_e (\ion{He}{i})}$ is $\lambda\lambda7281/6678$, because these lines are among the most intense  \ion{He}{i} lines in the optical range, their ratio is more sensitive to temperature but less sensitive to density than other \ion{He}{i} lines ratios, their recombination coefficients are more reliable than those for \ion{He}{i} $\lambda$4471 and $\lambda$5876, and their ratio is less affected by the effect of the interstellar extinction because of their close wavelengths.

Unfortunately, we could only measure the intensity of \ion{He}{i} $\lambda7281$ in two objects of our sample. In those objects where this line could not be measured we adopted as $\rm{T_e (\ion{He}{i})}$ the temperature derived from the line ratio $\lambda\lambda6678/5875$. 
We made this assumption because the determined $\rm{T_e (\ion{He}{i})}$ from $\lambda\lambda7281/6678$ and $\lambda\lambda6678/5875$ are very close, within the uncertainties, for Vy\,2-2 and IC\,4997; this result was also found by \citet{otsuka:10} for the halo PN BoBn\,1. 

\subsubsection{Temperatures and densities from \ion{O}{ii}}

It is known that the emissivities of recombination lines have a weak dependence on electronic temperature ($\rm{ I_{ORLs} \sim T_e^{-1}}$), nonetheless, some lines of \ion{O}{ii} emitted from states with different orbital angular momenta show some dependence on electronic temperature, thus they can be used to estimate the temperature in the zone where \ion{O}{ii} lines are being emitted. This methodology, first used by \citet{wesson:03} and then supported by some other authors \citep{tsamis:04,wesson:05,fang:13,mcnabb:13}, uses the intensity ratio $\lambda\lambda 4649/4089$, lines emitted in different transitions, as temperature indicator. The results presented by these works show that $\rm{T_e(\ion{O}{ii})}$ are very low compared to $\rm{T_e}$ from CELs, and follow a tendency of temperature values $\rm{ T_e([\ion{O}{iii}]) \geq T_e(BJ) \geq T_e(\ion{He}{i}) \geq T_e(\ion{O}{ii}) }$. The main disadvantage of this method is that \ion{O}{ii}$\lambda$4089 line is very faint and is difficult to measure. In this work, this $\ion{O}{ii}$ line could only be detected in the spectra of Hu\,2-1 and IC\,4997, therefore only for these two objects $\rm{T_e (\ion{O}{ii})}$ could be determined using this method.

Another way to estimate $\rm{T_e (\ion{O}{ii})}$ is by using the relationship proposed by \citet{peimbert:14}. This method is based on the rate of line intensities from both CELs and ORLs of O$^{+2}$ and relates their intensities with the electron temperature.

\begin{equation}
\rm{ \frac{I(V1)}{I(F1)} = 1.772 \, 10^{-5} \left( \frac{T_e}{10,000 \, K} \right)^{-0.40} \, exp \left( \frac{29,170 \, K}{T_e}\right) ,}
\label{eq:to2_peimbert}
\end{equation}

\noindent where $\rm{I(V1)}$ is the intensity of all the lines of the V1 multiplet of \ion{O}{ii} and $\rm{I(F1)}$ is the sum of the intensities of [\ion{O}{iii}] $\rm{\lambda 5007  \, {\rm and} \, \lambda 4959}$. V1 multiplet is constituted by the lines $\lambda \lambda$4638, 4642, 4649, 4651, 4661, 4673, 4676, 4696; and in LTE, the intensity of $\lambda$4649 represents the 39.7\% of the total multiplet intensity \citep{peimbert:14}. 

On the other side, the relative population of the ground-term fine-structure levels of some recombining ions, such as \ion{O}{ii} and \ion{N}{ii}, vary with the electron density. From the intensity ratio of two ORLs emitted in the same multiplet but from different parent levels, one can determine the electronic density \citep{fang:13}. These authors show that line ratios \ion{O}{ii} $\lambda \lambda 4649/4661$ and \ion{N}{ii} $\lambda \lambda 5679/5666$ are useful for the electronic density calculation. In our work we were able to determine the densities from \ion{O}{ii} $\lambda \lambda 4649/4661$ for Vy\,2-2, Vy\,1-2, Hu\,2-1 and IC\,4997.

All the results obtained from the methods described above are listed in Table \ref{table:recdensity-temperature}.

\defcitealias{peimbert:14}{P14}

\begin{table*}
    \centering
	\caption{Electron densities (in cm$^{-3}$) and temperatures (in K) derived from recombination lines}
    \begin{tabular}{lccccc}
    \hline
    & Cn\,3-1 & Vy\,2-2 & Hu\,2-1 & Vy\,1-2 & IC\,4997 \\ \hline \hline 
    $\rm{n_e}$(\ion{O}{ii}) 4649/4661 & --- & 19,900$\pm$4,300 & 31,600$\pm$18,000 & 7,900$\pm$4,300 & 5,000$\pm$2,500 \\ \hline
    $\rm{T_e}$(BJ) & 7,600$\pm$200 & --- & 7,700$\pm$400 & --- & 11,300$\pm$700 \\ \hline
    $\rm{T_e}$(\ion{He}{i}) 7281/6678 & --- & 8,300$\pm$300 & --- & --- & 8,900$_{-800}^{+1,100}$ \\
    $\rm{T_e}$(\ion{He}{i}) 7281/5875 & --- & 8,600$\pm$400 & --- & --- & 8,000$_{-900}^{+1,400}$ \\
    $\rm{T_e}$(\ion{He}{i}) 6678/5875 & --- & 7,600$\pm$500 & 6,900$\pm$1,500 & 8,800$\pm$3,200 & 10,500$\pm$1,100 \\
    $\rm{T_e}$(\ion{He}{i}) 6678/4471 & --- & 5,900$\pm$400 & 5,800$\pm$300 & 7,300$\pm$3,000 & 3,900$_{-300}^{+600}$ \\
    $\rm{T_e}$(\ion{He}{i}) 5875/4471 & 3,100$\pm$400 & 4,300$\pm$700 & 2,500$\pm$300 & 3,900$\pm$1,900 & 4,200$_{-600}^{+1,400}$ \\
    $\rm{T_e}$(\ion{He}{i}) \textit{Adopted$^1$} & 7,600$\pm$200 & 8,300$\pm$300 & 6,600$\pm$1,300 & 8,800$\pm$3,200 & 8,900$_{-800}^{+1,100}$ \\ \hline
    $\rm{T_e}$(\ion{O}{ii})   (\citetalias{peimbert:14}) & --- & 8,000$\pm$300 & 7,900$\pm$100 & 6,400$\pm$100 & 7,100$\pm$100 \\
    $\rm{T_e}$(\ion{O}{ii}) 4649/4089 & --- & --- & 2,900: & --- & 2,700: \\
    $\rm{T_e}$(\ion{O}{ii}) \textit{Adopted$^2$} & 7,600$\pm$200 & 8,000$\pm$300 & 7,900$\pm$100 & 6,400$\pm$100 & 7,100$\pm$100 \\ \hline
    \multicolumn{6}{l}{1: Adopted for the determination of He$^+$ abundances.} \\
    \multicolumn{6}{l}{2: Adopted for the determination of O$^{+2}$, N$^{+2}$ and C$^{+}$ abundances.} \\
    \multicolumn{6}{l}{: represents a very uncertain result.} \\
    \end{tabular}
    \label{table:recdensity-temperature}
\end{table*}

\section{Ionic abundances from CELs}

Ionic abundances for O$^+$, O$^{+2}$, N$^+$, Ne$^{+2}$, Ne$^{+3}$, S$^+$, S$^{+2}$, Cl$^{+2}$, Ar$^{+2}$, Ar$^{+3}$, Ar$^{+4}$, Fe$^{+}$, Fe$^{+2}$, and others, were determined from CELs using the task \textit{get.IonAbundance} from \textsc{pyneb}. For this we used the de-reddened intensities of the lines listed in Table \ref{table:lines_ionic} 
as measured for each nebula, and  the adopted physical conditions for each nebular zone as presented in Table \ref{table:cel_zones}. 
The results  for all the objects  are given in Table  \ref{table:abundances}.	In the following the adopted zones for each case are presented.

\subsection{Cn\,3-1}

For this object we assumed a single zone of temperature and density for all the ionic species. The adopted temperature is the mean value of the temperatures derived from [\ion{N}{ii}] and [\ion{O}{iii}] line ratios, which are equal within uncertainties.
The density was determined as the mean value of the computed densities from [\ion{S}{ii}], [\ion{O}{ii}] and [\ion{Cl}{iii}] line ratios. The density from [\ion{Fe}{iii}] which is higher, was not considered as the [\ion{Fe}{iii}] lines are very faint and the derived density has large uncertainty (see the $1\sigma$ error band in Fig. \ref{fig:diagnostics}). 
Thus, Cn\,3-1 is assumed to be a nebula with a single zone of temperature and density with values $\rm{T_e} \sim 7,900$ K and $\rm{n_e} \sim 6,600$ cm$^{-3}$.

\subsection{Vy\,2-2}
At least two density zones were found in this object: a low density zone given by the [\ion{S}{ii}] diagnostic with a value of $\rm{n_e} \sim 6,200$ cm$^{-3}$; and a high density zone given by the [\ion{Cl}{iii}] diagnostic with a value $\rm{n_e \sim 58,200}$ cm$^{-3}$, almost ten times larger than that of the low density zone. Most probably there is a gradient of density increasing inwards in this object. For the single ionized spices the density from [\ion{S}{ii}] ratio was used, while for the twice or more ionized spices the density from [\ion{Cl}{iii}] ratio was used.  Regarding the electron temperature, the [\ion{N}{ii}] $\lambda\lambda$(6548+6584)/5755 line ratio cannot be used as temperature diagnostic due to the nebular lines are highly affected by the inner high density, therefore these lines are undergoing collisional de-excitation. Thus  the [\ion{N}{ii}] ratio is more indicative of density, showing a value of about 10$^5$ cm$^{-3}$ (see Fig. \ref{fig:diagnostics}). The same occurs with the temperature sensitive [\ion{Ar}{iv}]$\lambda\lambda$7170/4740, which in this case is indicating a density of about 10$^6$ cm$^{-3}$ for the highest ionized zone.
Then the only available temperature diagnostics are the [\ion{O}{iii}] and [\ion{Ar}{iii}] line ratios providing temperatures of about 12,800 K and 8,800 K, respectively. We decided to adopt the temperature given by [\ion{Ar}{iii}] for the low ionized species, and $\rm{T_e}$([\ion{O}{iii}]) for the high ionized species, as indicated in Table \ref{table:cel_zones}.

\subsection{Hu\,2-1}
Densities were derived from [\ion{Cl}{iii}], [\ion{Fe}{iii}] and [\ion{Ar}{iv}] diagnostics, with values of 13,200 cm$^{-3}$, 32,300 cm$^{-3}$ and 48,700 cm$^{-3}$, respectively. The [\ion{O}{ii}] and [\ion{S}{ii}] density diagnostic ratios could not be calculated with \textsc{pyneb} routines due to apparent problems with atomic parameters. As it is shown in Fig. \ref{fig:diagnostics} the [\ion{S}{ii}] line ratio is useless for density determination. On the other hand, the [\ion{O}{ii}] density, previously determined by \citet{wesson:05} and \citet{delgado:15}, with a value of $\sim$7,000  cm$^{-3}$ can be adopted. Thus Hu\,2-1 is a nebula with increasing density inwards.

For the single and double ionized species we adopted the density given by the [\ion{Cl}{iii}] and for the triple ionized species we adopted the mean value of [\ion{Fe}{iii}] and [\ion{Ar}{iv}] densities, 47,100 cm$^{-3}$.
An electron  temperature of 10,200 K was determined from [\ion{N}{ii}] lines, and a value of 9,000 K from the [\ion{O}{iii}] lines in the zone were both diagnostics cross with [\ion{Ar}{iv}] density diagnostic. Therefore we assumed a single temperature for the whole nebula of 9,300 K, given by the mean value of [\ion{N}{ii}] and [\ion{O}{iii}] temperatures.

\subsection{Vy\,1-2}
In this case we adopted a single zone of density and temperature for all species. Density is given by the mean value of [\ion{S}{ii}], [\ion{O}{ii}], and [\ion{Ar}{iv}] diagnostics and leads to a value of 5,400 cm$^{-3}$. The slightly larger density provided by the [\ion{Cl}{iii}] lines was not considered, due to the large uncertainty this value presents (see the wide 1$\sigma$ pink error band in Fig. \ref{fig:diagnostics}). 
The temperature was determined as the mean value of [\ion{N}{ii}] and [\ion{O}{iii}] diagnostics which are very similar, leading a value of 9,500 K.

\subsection{IC\,4997}

This is a very complex object. \citet{hyung:94}, by analysing optical and ultraviolet spectra, concluded that this PN contains at least two important density zones: an outer zone with  $\rm{n_e} \sim 10^4$ cm$^{-3}$ and an inner one with $\rm{n_e \sim 10^6 - 10^7}$ cm$^{-3}$. One important detail about IC\,4997 is the low [\ion{O}{iii}] $\lambda\lambda4959/4363$ line ratio which shows a present value $\sim$ 2.05. This ratio has been decreasing with time \citep{arkhipova:17}. This low value is caused by the high intensity of the auroral line $\lambda4363$, augmented due to the large nebular density \citep{menzel:41}, and the low intensity of the nebular lines which, as a consequence of the density larger than the critical value for these lines, are undergoing collisional de-excitation. Therefore, it is more appropriate to use the ratio of the nebular to auroral line as a density diagnostic instead of a temperature one \citep[e.g.,][]{kingsburgh:94,wesson:05}.  Thus this ratio leads to a value of $\rm{n_e} \sim 1.81\times 10^6$ cm$^{-3}$. 

The diagnostic diagram for this object (Fig. \ref{fig:diagnostics}) is very complex. Our interpretation is that there are, at least, two very different density-zones: one given by [\ion{Cl}{iii}] and [\ion{Fe}{iii}] line ratios, of about 31,700 cm$^{-3}$, and a second one given by [\ion{O}{iii}] and [\ion{Ar}{iii}] line ratios which is larger than about 1.81$\times$10$^6$ cm$^{-3}$. The [\ion{Ar}{iv}] temperature diagnostic line ratio, density-sensitive for this case, also indicates such a high density. It was also possible to estimate the density given by [\ion{S}{ii}] lines but this diagnostic is in the limit of its validity range and the interpretation is inconclusive, therefore it is not presented in the diagram. 

The [\ion{N}{ii}] diagnostic ratio is the only one useful to estimate the electron temperature. The intersection between this ratio and the [\ion{Cl}{iii}] density diagnostics results in a $\rm{T_e = 16,200}$ K and $\rm{n_e = 31,700}$ cm$^{-3}$. We adopted only the density value of [\ion{Cl}{iii}] because of the high dispersion of [\ion{Fe}{iii}] density. When we assumed this $\rm{T_e}$ for the inner zones, we found $\rm{n_e([\ion{O}{iii}]) = 1.81 \times 10^6}$ cm$^{-3}$. 
To determine ionic abundances, we decided to adopt a single temperature zone and two different density zones, given by [\ion{Cl}{iii}] and [\ion{O}{iii}] ratios; they are shown in Table \ref{table:cel_zones}.

\begin{table}
 \centering
 \caption{Collisionally excitation lines used for ionic abundances determinations}
    \begin{tabular}{ll} \hline
    X$^{\rm{+i}}$ & Line\\ \hline \hline 
    N$^+$ & [\ion{N}{ii}] $\lambda\lambda$6548,6584 \\
    O$^+$ & [\ion{O}{ii}] $\lambda\lambda$3727,3729 \\
    O$^{+2}$ & [\ion{O}{iii}] $\lambda\lambda$4959,5007 \\
    Ne$^{+2}$ & [\ion{Ne}{iii}] $\lambda\lambda$3868,3967 \\
    Ne$^{+3}$ & [\ion{Ne}{iv}] $\lambda$4724 \\
    S$^+$ & [\ion{S}{ii}] $\lambda\lambda$6716,6731 \\
    S$^{+2}$ & [\ion{S}{iii}] $\lambda$6712 \\
    Cl$^{+2}$ & [\ion{Cl}{iii}] $\lambda\lambda$5517,5537 \\
    Ar$^{+2}$ & [\ion{Ar}{iii}] $\lambda\lambda$5192,7136 \\
    Ar$^{+3}$ & [\ion{Ar}{iv}] $\lambda\lambda$4711,4740 \\
    Ar$^{+4}$ & [\ion{Ar}{v}] $\lambda$7005 \\
    Fe$^{+}$ & [\ion{Fe}{ii}] $\lambda$7155 \\
    Fe$^{+2}$ & [\ion{Fe}{iii}] $\lambda\lambda$4659,4701,4734,4755 \\
    K$^{+3}$ & [\ion{K}{iv}] $\lambda$6102 \\
    \hline
    \end{tabular}
    \label{table:lines_ionic}
\end{table}

\section{Ionic abundances from ORLs: The ADFs}
Ionic abundances from ORLs were computed using \textsc{pyneb} \textit{get.IonAbundance} with the previously determined temperatures and densities ($\rm{T_e}$(\ion{He}{i}), $\rm{T_e}$ (\ion{O}{ii}) or $\rm{T_e}$ (BJ)) presented in Table \ref{table:recdensity-temperature}. We adopted the density derived from \ion{O}{ii} $\lambda\lambda 4661/4649$ to determine ionic abundances from recombination lines, except for Cn\,3-1 where it was not possible to estimate it and thus the density determined from forbidden lines was adopted. Adopted temperatures were different in each case: for \ion{He}{i} we adopted the derived $\rm{T_e}$(\ion{He}{i}), for \ion{He}{ii} in Vy\,1-2, the temperature determined from CELs, and for heavy elements, $\rm{T_e}$ (\ion{O}{ii}) derived from \citet{peimbert:14} expression. We decided not to use $\rm{T_e}$ (\ion{O}{ii}) determined in Hu\,2-1 and in IC\,4997 from line ratio $\lambda \lambda 4649/4089$ due to their very large uncertainties. 
For Cn\,3-1 we used $\rm{T_e}$ (BJ) to determine the abundances of all recombination lines.

For each ion, different line intensities were used: $\lambda$5875 for \ion{He}{i},  $\lambda$4686 for \ion{He}{ii}, $\lambda$4267 for \ion{C}{ii}, the intensities of V1 multiplet for \ion{O}{ii} and the intensities of multiplet V3 for \ion{N}{ii}.

The results for O$^{+2}$, C$^{+2}$, and N$^{+2}$ are presented in Table \ref{orls_abundances}.
The O$^{+2}$ values, compared to the ionic abundances from CELs, for the same ion, allow us to derive the ADFs that are presented in Table \ref{adfs_table}.

\begin{table*}
\centering
\caption{\bf Ionic abundances from ORLs}
\begin{tabular}{lcccccc} \hline
Multiplet & $\lambda_0$ & Cn\,3-1 & Vy\,2-2 & Hu\,2-1 & Vy\,1-2 & IC\,4997 \\ \hline \hline

\multicolumn{7}{c}{$\rm{He^{+}/H^+}$ } \\ \hline
 & 5875 & 0.054$\pm$0.002 & 0.115$\pm$0.001 & 0.107$\pm$0.004 & 0.900$\pm$0.008 & 0.146$\pm$0.005 \\ 
 \hline \hline
 
\multicolumn{7}{c}{$\rm{He^{+2}/H^+}$ } \\ \hline  
 & 4686 & --- & --- & --- & 0.030$\pm$0.001 & --- \\
 \hline \hline 

\multicolumn{7}{c}{$\rm{O^{+2}/H^+ (\times 10^{-4})}$} \\ \hline
V1  & 4638 & --- & 5.09$\pm$1.42 & 4.50$\pm$0.65 & ---             & 7.25$\pm$2.20 \\
    & 4641 & --- & 3.83$\pm$0.41 & 5.06$\pm$0.42 & 40.90$\pm$8.27  & 8.47$\pm$1.12 \\
    & 4649 & --- & 6.47$\pm$1.30 & 3.45$\pm$0.30 & 28.54$_{-1.59}^{3.23}$  & 8.25$\pm$1.20 \\
    & 4650 & --- & 9.24$_{-5.79}^{+4.78}$ & ---  & 18.22$\pm$8.10  & 7.76$\pm$2.05 \\
    & 4661 & --- & 6.41$\pm$0.93 & 3.44$_{-0.65}^{+0.53}$ & 26.81$\pm$8.75  & 7.31$\pm$1.09 \\
    & 4673 & --- & ---           & ---           & 81.13$_{-47.71}^{+58.58}$ & --- \\
    & 4676 & --- & ---           & 3.92$\pm$0.71 & ---             & 6.68$\pm$1.58 \\
    & 4696 & --- & ---           & ---           & ---             & --- \\
    & \textit{V1 sum} & --- & 5.50$\pm$0.71 & 3.93$\pm$0.19 & 30.95$\pm$2.90 & 8.54$\pm$0.61 \\
 \hline \hline
 
\multicolumn{7}{c}{$\rm{C^{+2}/H^+ (\times 10^{-4})}$} \\ \hline
6 & 4267 & 1.15$\pm$0.27 & 3.00$\pm$0.55 & 3.70$\pm$0.20 & 8.56$\pm$2.60 & 1.81$\pm$0.34 \\
\hline \hline

\multicolumn{7}{c}{$\rm{N^{+2}/H^+ (\times 10^{-4})}$} \\ 
\hline

V3  & 5667 & ---    & 3.39$\pm$0.61 & 1.90$\pm$0.38 & ---           & 2.77$\pm$0.74 \\
    & 5676 & ---           & ---    & 2.11$\pm$0.75 & ---           & --- \\
    & 5679 & ---           & ---    & 1.38$\pm$0.21 & 3.55$\pm$1.78 & 1.14$\pm$0.51 \\
    & 5711 & 2.09$\pm$0.95 & ---    & ---           & ---           & --- \\
    & \textit{V3 sum} & 1.99$\pm$0.90 & 4.18$\pm$0.70 & 1.73$\pm$0.20 & 3.40$\pm$1.68 & 1.77$\pm$0.57 \\
 \hline
\end{tabular}
\label{orls_abundances}
\end{table*}

\section{Abundance results}
Total abundances from CELs were calculated from the measured ionic abundances and using Ionization Correction Factors,  ICFs, to correct for the unseen ions. The ICFs employed and their expressions are presented in Appendix \ref{section:icfs_exp}. Values for ICFs are listed in Table \ref{table:abundances} together with the total abundances derived for each object.

In \S8 we discuss the derived abundances, together with the kinematics, for each case.

\begin{table*}
\caption{\bf Ionic and total abundances from CELs. He/H total abundances from ORLs are also presented.}
\begin{tabular}{lccccc} \hline
Ion & Cn\,3-1 & Vy\,2-2 & Hu\,2-1 & Vy\,1-2 & IC\,4997 \\ \hline \hline 
O$^{+} (\times 10^{-5})$ & 37.10$_{-6.67}^{+11.92}$ & 1.39$_{-0.29}^{+0.43}$ & 5.70$_{-5.70}^{+8.89}$ & 1.57$_{-0.95}^{+1.47}$ & 0.72$_{-0.38}^{+1.23}$ \\
O$^{+2} (\times 10^{-4})$ & 0.22$\pm$0.02 & 1.22$_{-0.13}^{+0.51}$ & 2.12$_{-0.76}^{+2.84}$ & 5.65$_{-0.95}^{+1.47}$ & 1.72$_{-0.79}^{+1.99}$ \\
ICF(O) & 1.00 & 1.00 & 1.00 & 1.17$\pm$0.02 & 1.00 \\ \hline
N$^{+} (\times 10^{-5})$ & 6.24$_{-0.59}^{+0.80}$ & 0.55$\pm$0.05 & 1.86$_{-0.48}^{+1.38}$ & 0.66$_{-0.11}^{+0.17}$ & 0.17$_{-0.06}^{+0.12}$ \\
ICF(N) & 1.06$\pm$0.01 & 10.28$_{-2.45}^{+3.74}$ & 3.99$_{-0.55}^{+0.70}$ & 44.45$_{-13.34}^{+10.35}$ & 24.37$_{-5.55}^{+6.54}$ \\ \hline
Ne$^{+2} (\times 10^{-5})$ & 0.03$\pm$0.01 & 3.34$_{-0.34}^{+1.16}$ & 2.54$_{-0.95}^{+3.53}$ & 9.84$_{-2.09}^{+2.73}$ & 3.78$_{-1.47}^{+3.03}$ \\
Ne$^{+3} (\times 10^{-4})$ & --- & --- & --- & 3.82$_{-1.98}^{+3.88}$ & --- \\
ICF(Ne) & 17.45$_{-2.20}^{+3.71}$ & 1.11$\pm$0.04 & 1.33$\pm$0.07 & 1.20$\pm$0.02 & 1.04$\pm$0.01\\ \hline
Ar$^{+2} (\times 10^{-6})$ & --- & 2.73$_{-0.15}^{+0.22}$ & 1.10$_{-0.45}^{+1.97}$ & --- & 0.47$_{-0.15}^{+0.33}$ \\
Ar$^{+3} (\times 10^{-7})$ & --- & 0.16$_{-0.03}^{+0.06}$ & 0.19$_{-0.06}^{+0.25}$ & 14.79$_{-2.77}^{+4.22}$ & 1.91$_{-0.87}^{+2.30}$ \\
Ar$^{+4} (\times 10^{-8})$ & --- & --- & --- & 1.34$_{-0.36}^{+0.40}$ & --- \\
ICF(Ar) & --- & 1.11$\pm$0.04 & 1.33$\pm$0.08 & 1.02$\pm$0.01 & 1.04$\pm$0.01 \\ \hline
S$^{+} (\times 10^{-7})$ & 17.06$_{-2.52}^{+4.70}$ & 0.51$_{-0.08}^{+0.14}$ & 1.58$_{-0.45}^{+1.02}$ & 3.97$_{-1.01}^{+2.30}$ & 1.81$_{-0.78}^{+2.34}$ \\
S$^{+2} (\times 10^{-6})$ & 4.12$_{-0.42}^{+0.62}$ & 12.76$_{-1.38}^{+1.58}$ & 1.74$_{-0.60}^{+2.51}$ & 3.85$_{-0.92}^{+1.31}$ & 1.17$_{-0.42}^{+0.81}$ \\
ICF(S) & 1.00$\pm$0.01 & 1.56$\pm$0.14 & 1.20$\pm$0.05 & 2.47$_{-0.27}^{+0.17}$ & 2.04$\pm$0.16 \\ \hline
Cl$^{+2} (\times 10^{-8})$ & 9.03$_{-0.97}^{+1.38}$ & 6.17$_{-1.27}^{+3.41}$ & 3.78$_{-1.04}^{+3.53}$ & 9.40$_{-2.29}^{+3.28}$ & 1.09$_{-0.38}^{+0.81}$ \\
ICF(Cl) & 1.41$\pm$0.08 & 1.57$\pm$0.14 & 1.31$\pm$0.05 & 2.73$\pm$0.19 & 2.37$\pm$0.10 \\ \hline
Fe$^{+} (\times 10^{-7})$ & --- & 3.65$_{-0.34}^{+0.44}$ & --- & --- & --- \\
Fe$^{+2} (\times 10^{-7})$ & 3.84$\pm$0.48 & 7.40$\pm$0.88 & 1.85$_{-0.60}^{+2.23}$ & --- & 0.48$_{-0.13}^{+0.27}$ \\
ICF(Fe) & --- & 0.31$\pm$0.06 & 0.59$\pm$0.05 & --- & --- \\ \hline
K$^{+3} (\times 10^{-9})$ & --- & --- & --- & 16.53$_{-6.86}^{+7.46}$ & 1.16$_{-0.48}^{+0.77}$ \\ 
ICF(K) & --- & --- & --- & 1.53$\pm$0.02 & --- \\ \hline
\hline
\multicolumn{5}{l}{Total abundances in a 12+log(X/H) scale}\\
\hline
He/H & 10.73$\pm$0.02 & 11.06$\pm$0.01 & 11.03$\pm$0.02 & 11.07$\pm$0.03 & 11.16$\pm$0.01 \\
O/H & 8.60$_{-0.08}^{+0.12}$ & 8.14$_{-0.05}^{+0.12}$ & 8.46$_{-0.20}^{+0.38}$ & 8.84$_{-0.08}^{+0.10}$ & 8.25$_{-0.26}^{+0.34}$ \\
N/H & 7.82$\pm$0.05 & 7.74$_{-0.09}^{+0.13}$ & 7.87$_{-0.12}^{+0.22}$ & 8.46$_{-0.11}^{+0.10}$ & 7.63$_{-0.15}^{+0.20}$ \\
Ne/H & 6.79$\pm$0.16 & 7.58$_{-0.05}^{+0.11}$ & 7.52$_{-0.14}^{+0.38}$ & 8.07$_{-0.10}^{+0.11}$ & 7.60$_{-0.21}^{+0.26}$ \\
Ar/H & --- & 6.48$\pm$0.03 & 6.18$_{-0.23}^{+0.46}$ & 6.22$_{-0.09}^{+0.11}$ & 5.84$_{-0.20}^{+0.26}$ \\
S/H & 6.77$_{-0.05}^{+0.07}$ & 7.31$\pm$0.05 & 6.37$_{-0.18}^{+0.36}$ & 7.02$_{-0.10}^{+0.12}$ & 6.44$_{-0.18}^{+0.23}$ \\
Cl/H & 5.11$_{-0.06}^{+0.08}$ & 4.97$_{-0.11}^{+0.23}$ & 4.70$_{-0.13}^{+0.27}$ & 5.41$\pm$0.11 & 4.41$_{-0.18}^{+0.23}$ \\
Fe/H & --- & 6.35$_{-0.06}^{+0.08}$ & 5.63$_{-0.16}^{+0.35}$ & --- & --- \\
K/H & --- & --- & --- & 4.45$_{-0.24}^{+0.16}$ & --- \\
\hline
N/O & $-$0.78$_{-0.09}^{+0.06}$ & $-$0.41$_{-0.08}^{+0.07}$ & $-$0.60$_{-0.18}^{+0.11}$ & $-$0.38$_{-0.15}^{+0.10}$ & $-$0.62$_{-0.20}^{+0.15}$ \\
Ne/O & $-$1.82$_{-0.11}^{+0.10}$ & $-$0.56$\pm$0.02 & $-$0.92$\pm$0.03 & $-$0.76$\pm$0.03 & $-$0.66$_{-0.08}^{+0.06}$ \\
Ar/O & --- & $-$1.65$_{-0.15}^{+0.06}$ & $-$2.26$_{-0.07}^{+0.11}$ & $-$2.62$\pm$0.03 & $-$2.42$_{-0.09}^{+0.07}$ \\
S/O & $-$1.82$_{-0.06}^{+0.04}$ & $-$0.85$_{-0.13}^{+0.07}$ & $-$2.08$_{-0.05}^{+0.04}$ & $-$1.82$_{-0.07}^{+0.06}$ & $-$1.81$\pm$0.52 \\
Cl/O & $-$3.49$\pm$0.04 & $-$3.16$_{-0.08}^{+0.11}$ & $-$3.75$_{-0.12}^{+0.07}$ & $-$3.44$\pm$0.09 & $-$3.84$_{-0.36}^{+0.25}$ \\
Fe/O & --- & $-$1.80$_{-0.10}^{+0.07}$ & $-$2.81$\pm$0.05 & --- & --- \\
K/O & --- & --- & --- & $-$4.39$_{-0.21}^{+0.16}$ & --- \\ 
\hline
\end{tabular}
\label{table:abundances}
\end{table*}

\begin{table*}
\centering
\caption{\bf Abundance Discrepancy Factors (ADFs) of the sample}
\begin{tabular}{lccccc} \hline
 & Cn\,3-1 & Vy\,2-2 & Hu\,2-1 & Vy\,1-2 & IC\,4997 \\ \hline \hline 
$\rm{O^{+2}/H^{+}} \times10^{-4}$   (ORLs) & --- & 5.50$\pm$0.71 & 3.93$\pm$0.19 & 30.95$\pm$2.90 & 8.54$\pm$0.61 \\
$\rm{O^{+2}/H^{+}} \times10^{-4}$   (CELs) & 0.22$\pm$0.02 & 1.22$_{-0.13}^{+0.51}$ & 2.12$_{-0.76}^{+2.84}$ & 5.65$_{-0.95}^{+1.47}$ & 1.72$_{-0.79}^{+1.99}$\\
ADF(O$^{+2}$) & --- & 4.30$_{-1.16}^{+1.00}$ & 1.85$\pm$1.05 & 5.34$_{-1.08}^{+1.27}$ & 4.87$_{-2.71}^{+4.34}$ \\ \hline
\end{tabular}
\label{adfs_table}
\end{table*}

\section{The kinematics}

\begin{figure*}
	\includegraphics[width=\columnwidth]{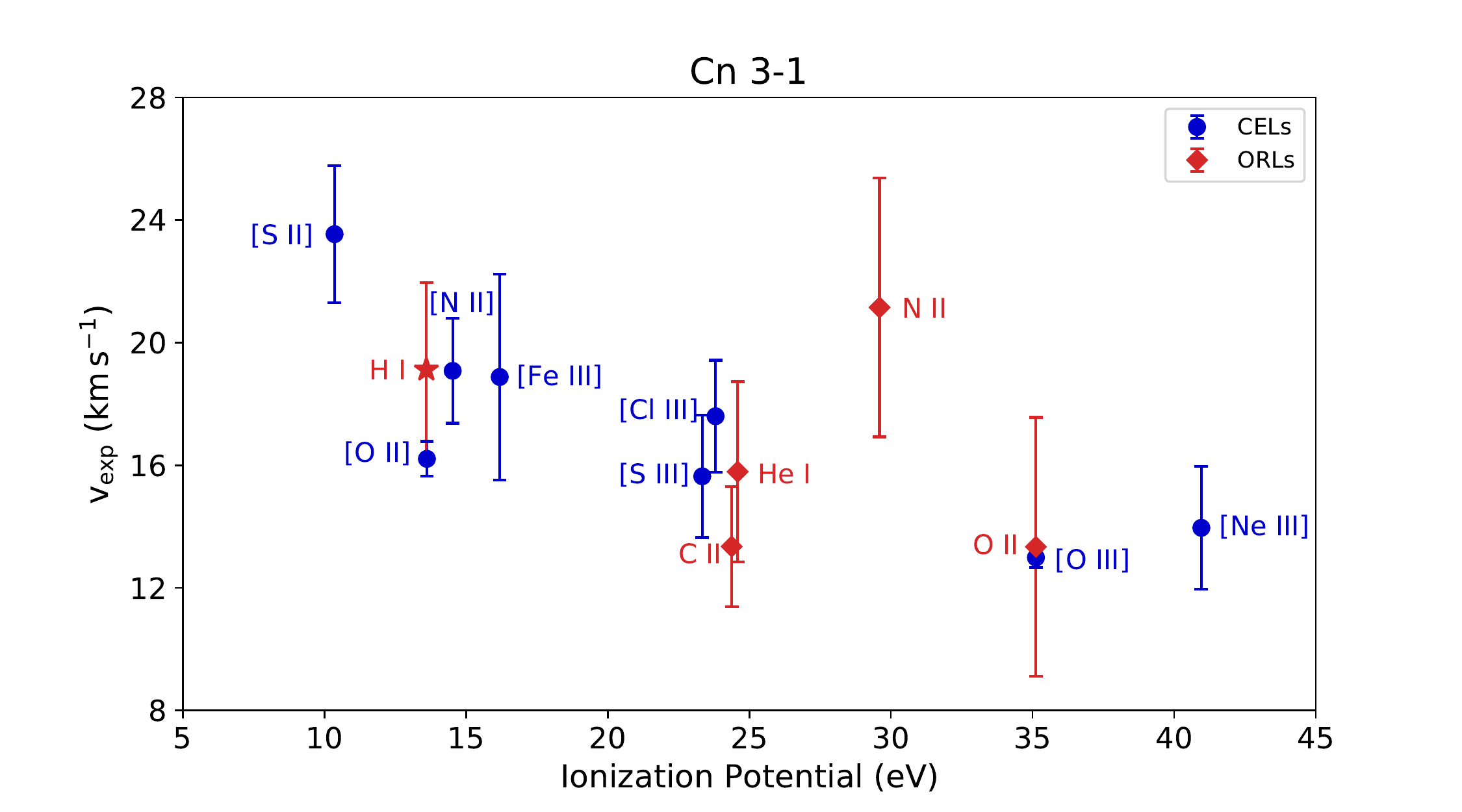}
	\includegraphics[width=\columnwidth]{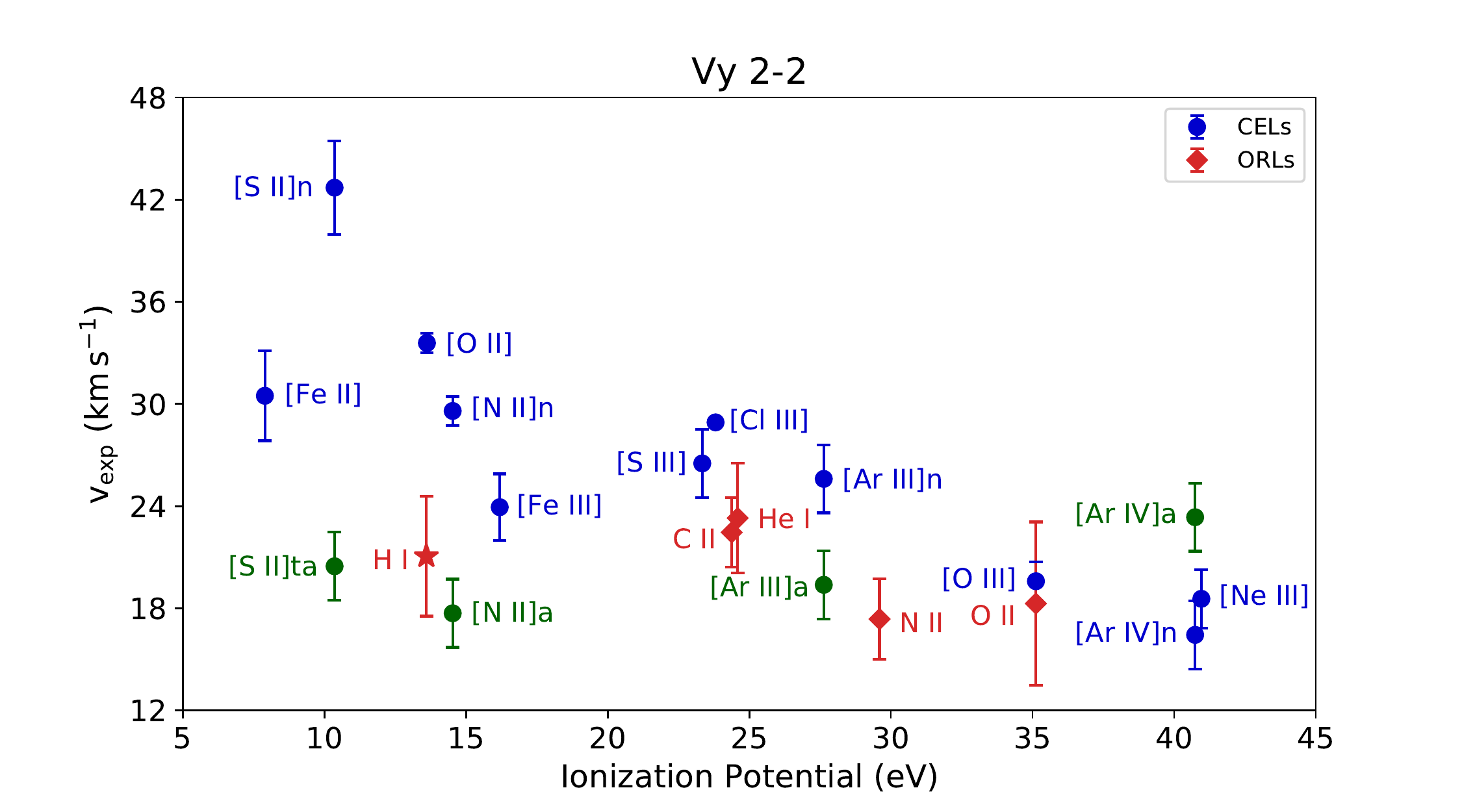}
	\includegraphics[width=\columnwidth]{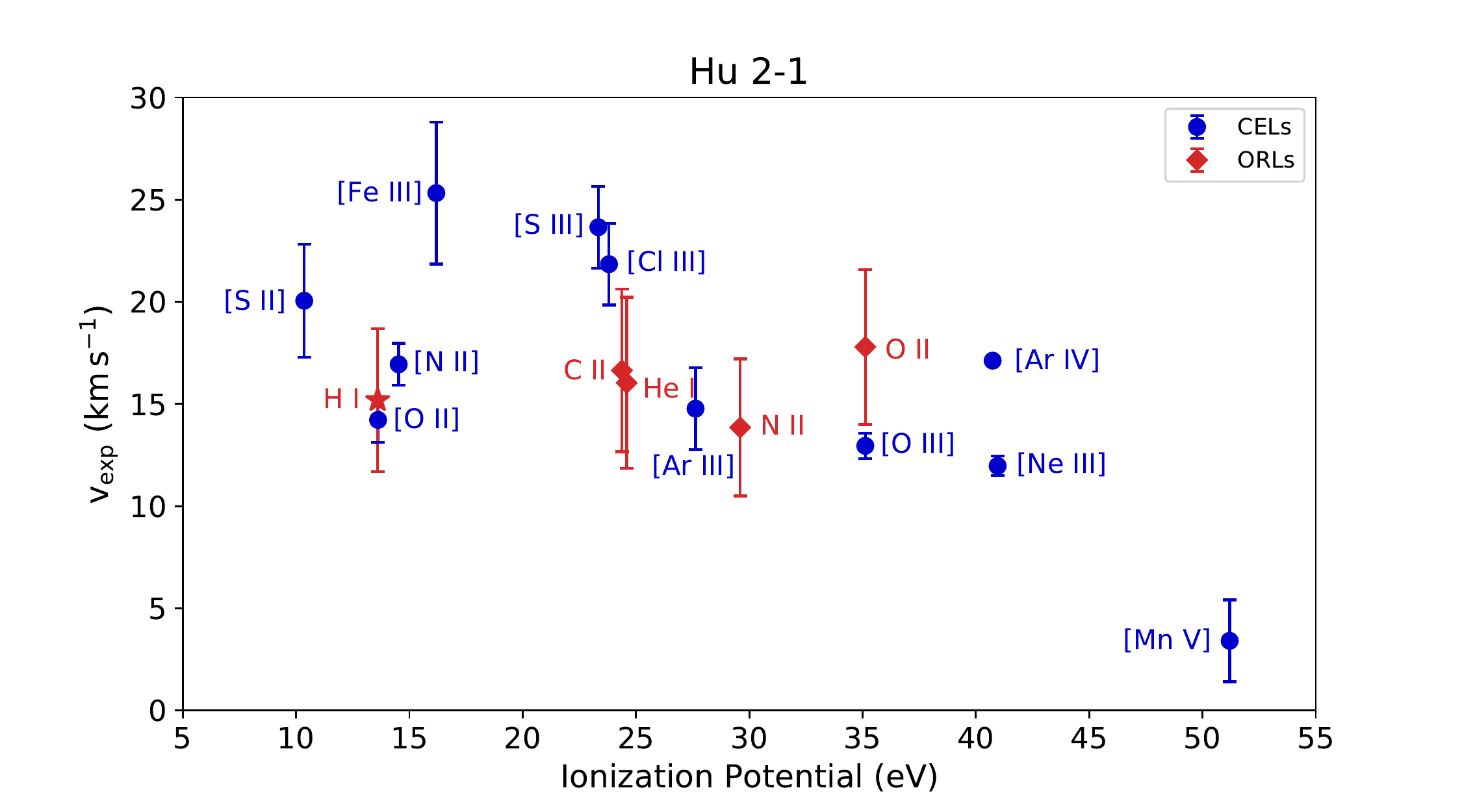}
	\includegraphics[width=\columnwidth]{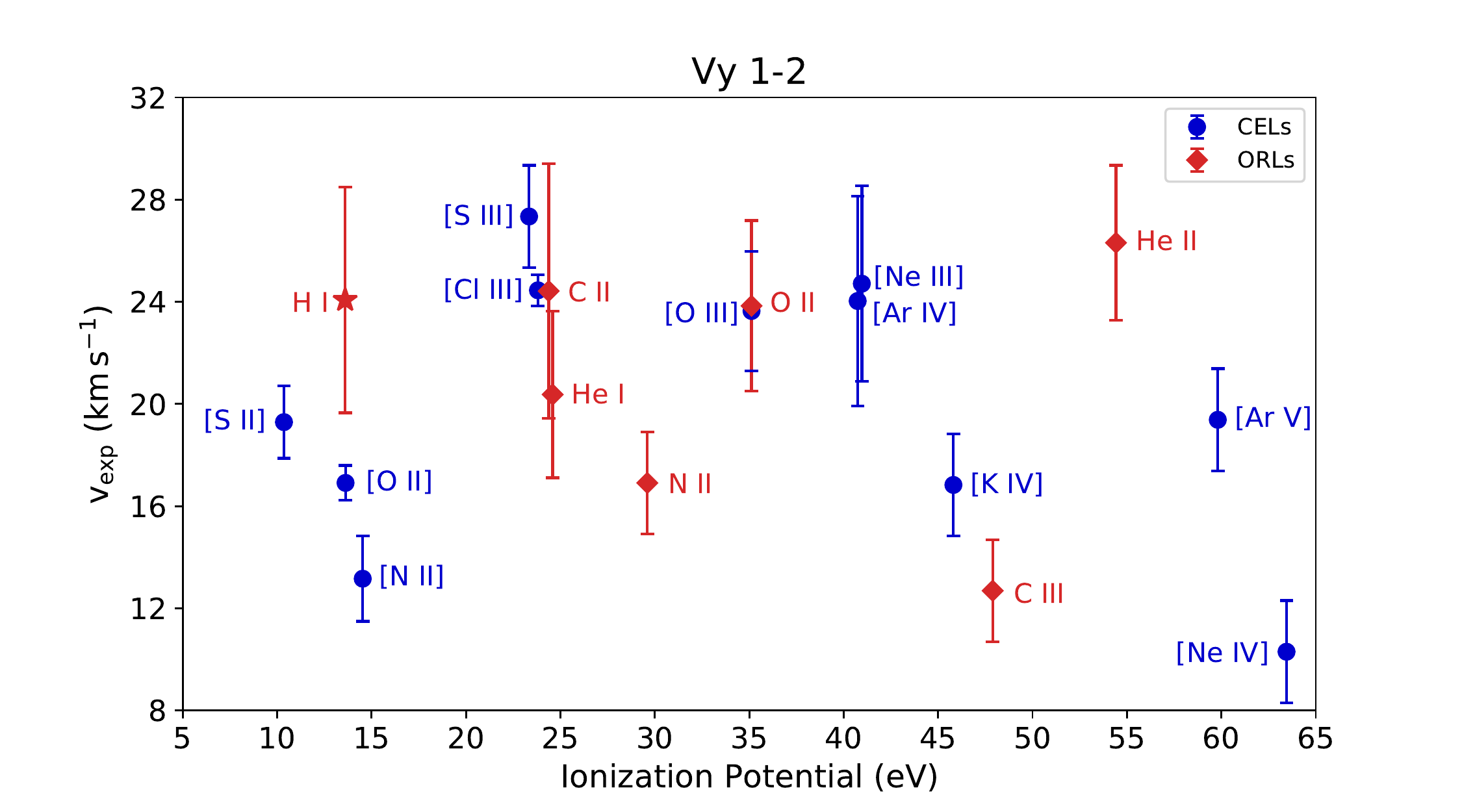}
	\includegraphics[width=\columnwidth]{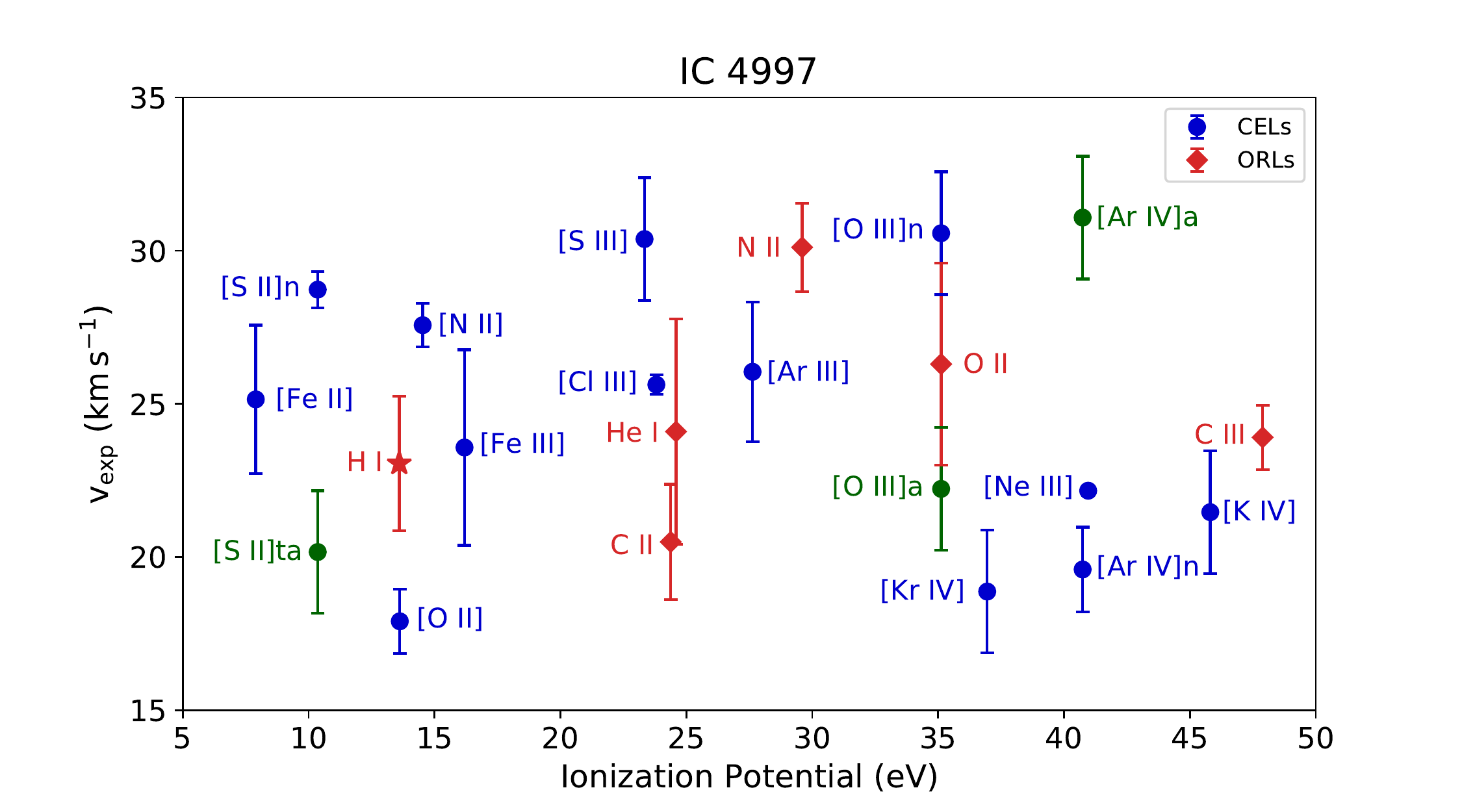}

    \caption{Expansion velocity vs. ionization potential of different ions, for the objects studied here. Expansion velocities derived from CELs are in blue, expansion velocities from ORLs are in red. In the cases of Vy\,2-2 and IC\,4997, the nebular and auroral lines of some ions are marked in green as they show different expansion velocities.}
    \label{fig:kinematics}
\end{figure*}

The expansion velocity of the different ions as a function of their ionization potential has been analysed. Due to the ionization structure of the nebulae, the ionization potential approximately represents the distance of the ion to the central star. Fig. \ref{fig:kinematics} shows these behaviours, including the expansion velocities given by CELs (in blue) and by ORLs (in red). Each dot represents the average value of the expansion velocities of the observed lines of each ion, while the error bars represent their mean standard deviation. When only a single line was available, we adopted $\rm{2 \, km \, s^{-1}}$ as the error value. Even when we detected in our spectra some permitted lines of \ion{O}{i}, \ion{Si}{ii}, \ion{N}{iii} and \ion{O}{iii} (Fig. \ref{fig:spectra}), these lines are emitted by different mechanisms than recombination, such as fluorescence and Bowen mechanism, therefore those lines were not considered for the comparison between CELs and ORLs.

Expansion velocities ($\rm{V_{exp}}$) were determined from the FWHM (Full Width at Half Maximum) for most of the lines which have a single component. 
As said in \S 2.1 the instrumental and thermal widths were discounted by assuming they add in quadrature.
 Turbulent velocities are  contributing also to the FWHM. Such velocities have been analysed, among others, by \citet{sabbadin:08, gesicki:03b, gesicki:03a}. The latter authors concluded  that turbulence is significant in PNe with [WC] central stars, but has small values in other PNe.
In this paper we are analysing non-[WR] nebula, so turbulence might not be important and we have not considered it in the expansion velocity calculation.

For the lines with a double-peaked profile, a fit of two deblending Gaussian was applied, and the expansion velocities were determined from the difference in velocity of both Gaussian.

Final expansion velocity of each ion was determined as the average value of the expansion velocities of their observed CELs and ORLs. Expansion velocities vs. ionization potentials are presented in Fig. \ref{fig:kinematics}.

In most cases we found that the kinematic behaviour of CELs is typical of a plasma expanding in the vacuum. That is, the expansion velocity increases with the distance to the central star, which is normally predicted for a plasma expanding with velocities style Hubble-flow.

The analysis of Fig. \ref{fig:kinematics} for the different objects is presented in the next section where a brief description of the objects is included, and we discuss the abundance results and the kinematics obtained.

Radial velocities ($\rm{V_{rad}}$) were determined for each observed line by using Doppler effect. The heliocentric correction was applied to all the observed velocities and they were averaged to get a systemic velocity for each object. These values and their standard deviations are presented in Table \ref{tab:radial_vels}.

\begin{table}
\centering
\caption{Heliocentric radial velocities for the sample of PNe.} 
    \begin{tabular}{lc}
    \hline
    Object   & $\rm{V_{rad}}$ \\ \hline \hline 
    Cn\,3-1  & $+29.62\pm6.37$  \\
    Vy\,2-2  & $-63.91\pm6.60$ \\
    Hu\,2-1  & $+23.62\pm5.67$ \\
    Vy\,1-2  & $-82.40\pm7.87$ \\
    IC\,4997 & $-58.36\pm9.91$ \\ \hline
    \end{tabular}
    \label{tab:radial_vels}
\end{table}

\section{Discussion of individual results}

\subsection{Cn\,3-1}
This is a dense and compact PN, with a low ionization degree corroborated by the low [\ion{O}{iii}] $\lambda$5007/H$\beta$ flux ratio of 0.28, therefore almost all the oxygen is as O$^+$. The nebula shows a low expansion velocity of about 12 km s$^{-1}$ from the [\ion{O}{iii}] $\lambda$5007 line (see Fig. \ref{fig:kinematics}), and a heliocentric velocity of $+$29.62$\pm$6.37 km s$^{-1}$ (Table \ref{tab:radial_vels}). 

Its distance from GAIA DR2 (7.65 kpc) and morphology from \citet{sahai:11} were already presented in Table \ref{tab:PNe}. 

Physical conditions and chemical abundances were derived previously for this nebula by  \citet{milingo:10} and \citet{wesson:05}. Our values are very similar to those of both authors (see Table \ref{tab:comparison_abundances}). The O abundance of $\rm{12+log(O/H) = 8.60}$, together with the heliocentric velocity, indicate that this is a disc nebula, located in the solar vicinity. As a new result we have determined the Ne abundance in this object, with a value $\rm{ log(Ne/O) \sim -1.82}$, which results to be very low in comparison with what is expected for normal PNe. There are only a handful of PNe with such a low Ne/O abundance ratio. The value reported for Cn\,3-1 is similar to the value reported for the halo PN H\,4-1 \citep{otsuka:13}, and for the PNe M\,1-14,  IC\,418 and  Mz3 \citep{henry:10,stanghellini:10}. No reliable explanation has been offered so far for this phenomenon. 

No ADF has been derived for this nebula previously. In this work we could not derive an ADF either, because the O$^{+2}$ abundance from ORLs could not be determined due to the low ionization degree of the nebula and due to the presence of stellar emission lines in the wavelength range where the most important \ion{O}{ii} lines (multiplet V1) are found.

Regarding the kinematic behaviour, the blue points in Fig. \ref{fig:kinematics} (up, left) show that low ionization CELs from [\ion{S}{ii}], [\ion{N}{ii}], and  [\ion{O}{ii}] present higher expansion velocities (from 18 to 21 km s$^{-1}$) than higher ionization CELs showing $\rm{V_{exp}}$ lower than 15 km s$^{-1}$. This behaviour is typical of an ionized plasma expanding in vacuum, accelerating with the distance to the central star.
The red dots in this figure, showing the expansion velocities given by the ORLs, present a similar behaviour to that of CELs (increasing velocity with distance to the central star), except for \ion{N}{ii} which shows a too large velocity for its ionization potential. However, considering the large error bars, \ion{N}{ii} velocity coincides with ions of similar ionization potential. It should be noticed that the \ion{O}{ii} recombination lines and [\ion{O}{iii}] collisionally excited lines present a very similar expansion velocity. Therefore, from these results there is no evidence of different plasmas with different kinematics producing recombination lines and collisionally excited lines in Cn\,3-1.

\subsection{Vy\,2-2}
This is a very compact, young and dense nebula.
It shows a heliocentric radial velocity of $-63.91 \pm 6.60$ km s$^{-1}$ (this work).
The distance from \citet{frew:16}, the morphology from  \citet{sahai:11}, the central star classification and the previous reported ADFs, as well as the huge FWZI at the base of H$\alpha$ were presented in Table \ref{tab:PNe}.
	
The \ion{He}{ii} $\lambda$4686 recombination line was detected here showing a wide profile, therefore we consider it of stellar origin.
Our diagnostic diagram indicates an external zone with density of $\rm{n_e([\ion{S}{ii}]) = 6,200}$ cm$^{-3}$, and  an inner zone of very high density with values $\rm{n_e}$([\ion{Cl}{iii}]) of about 58,200 cm$^{-3}$, while the [\ion{N}{ii}] and the [\ion{Ar}{iv}] temperature sensitive diagnostic lines are indicating densities of about $10^5 - 10^6$ cm$^{-3}$. Therefore a steep density gradient exists in this nebula. Electron temperatures could be derived from [\ion{O}{iii}] and [\ion{Ar}{iii}] CELs providing values of  $\rm{T_e}$([\ion{O}{iii}]) = 12,800 K and  $\rm{T_e}$([\ion{Ar}{iii}]) = 8,800 K. The high electron density prevents temperature determination from the [\ion{N}{ii}] $\lambda\lambda$(6548+6584)/5755 lines ratio.

The abundances derived for this nebula indicate a low-abundance object with $\rm{12+log(O/H) = 8.14}$, while it presents about solar N, Ne, S, Ar, and Cl abundances. The abundances found in the literature  by \citet{perinotto:04} and \citet{wesson:05} are in general agreement with our values (Table \ref{tab:comparison_abundances}). The low O abundance could be due to depletion in dust grains,  and since the S abundance is not largely depleted, the dominant dust in Vy\,2-2 might be the one of dual-chemistry (DC) \citep{garcia-hernandez:14}. We also found a $\rm{12+log(Fe/H) = 6.35_{-0.06}^{+0.08}}$, which is lower than the solar value of 7.46, this can be also an indicator of depletion into dust grains \citep[and references therein]{garcia-rojas:13}. The ADF(O$^{+2}$) derived in this work for O$^{+2}$ is $4.30_{-1.16}^{+1.00}$, which is significantly lower than the value 11.8 reported by \citet{wesson:05}.

The expansion velocities of the different ions, derived from the FWHM of lines, are presented in Fig. \ref{fig:kinematics} (up,right). It is found that CELs (blue dots) show a velocity field increasing with the distance from the central star, from about 19 km s$^{-1}$ in the  Ar$^{+3}$ and Ne$^{+2}$ zones to about 34 km s$^{-1}$ or larger in the zone where O$^+$, S$^+$ and Fe$^{+2}$ reside, while ORLs present slightly lower  $\rm{V_{exp}}$, and no clear gradient is found. The [\ion{O}{iii}] lines show $\rm{V_{exp} = 18}$ km s$^{-1}$, while   the \ion{O}{ii} lines show $\rm{V_{exp} = 15}$ km s$^{-1}$, which can be considered equal within uncertainties. 

An interesting issue occurring in this nebula, and also in IC\,4997, is that the auroral and trans-auroral lines, shown in green in Fig. \ref{fig:kinematics} (up, right) (e.g., [\ion{N}{ii}]$\lambda$5755, [\ion{S}{ii}]$\lambda$4068, [\ion{Ar}{iii}]$\lambda$5192) present lower expansion velocities than the nebular lines of the same ion ([\ion{N}{ii}]$\lambda \lambda6584, 6548$, [\ion{S}{ii}]$\lambda \lambda6716, 6731$, [\ion{Ar}{iii}]$\lambda$7135). The differences in velocity are larger than the uncertainties.  Exceptions are the lines of [\ion{Ar}{iv}] which show the opposite, the auroral line $\lambda$7170 presents slightly larger expansion velocity than the nebular line $\lambda$4740, however both velocities are similar within uncertainties. A possible explanation for this peculiar phenomenon is associated with the very high density gradient in this nebula. It is known that auroral lines usually have critical densities larger than the nebular lines thus, in very high density nebulae the auroral lines can be emitted in denser zones nearer the central star, where nebular lines are undergoing collisional de-excitation. These inner zones are expanding more slowly, in a Hubble-type flow.
 A more detailed discussion on this subject can be found in \S8.6.

\subsection{ Hu\,2-1}
Chemical abundances for this object have been reported by \citet{henry:10}, \citet{perinotto:04}, \citet{wesson:05} and \citet{delgado:15}. Their results are in agreement with the values reported here (Table \ref{tab:comparison_abundances}). With a heliocentric velocity of $+$23.62$\pm$5.67 km s$^{-1}$ and $\rm{12+log(O/H) = 8.46}$, Hu\,2-1 is a regular disc PN in the Galaxy.
\citet{wesson:05} reported an ADF(O$^{+2}$) of 4.00, while we have found an ADF(O$^{+2}$) of 1.85$\pm$1.05.  A discussion on this and other  differences is presented in \S9. 

This and Vy\,2-2 are the only objects in the sample in which we can determine an iron total abundance using the ICF given by \citet{rodriguez:05}, in the other objects their parameters are outside the ICF validity range. We found a value of $\rm{12+log(Fe/H) = 5.63}$, which is much lower than the solar value of $\rm{12+log(Fe/H) = 7.46}$ and, as for Vy\,2-2, may be indicative of iron depletion into dust \citep[and references therein]{garcia-rojas:13}, although O and S do not appear very depleted.

The kinematic behaviour of CELs (blue points in Fig. \ref{fig:kinematics} (middle,left)) is consistent with a Hubble-flow type expansion, as $\rm{V_{exp}}$ of [\ion{N}{ii}], [\ion{S}{ii}], [\ion{O}{ii}] and [\ion{S}{iii}] lines are larger than  $\rm{V_{exp}}$ of higher ionized species. $\rm{V_{exp}}$ of ORLs (red points) seem to show a flatter gradient. The velocity of \ion{O}{ii} ORLs appears slightly higher than $\rm{V_{exp}}$ of [\ion{O}{iii}] CELs but equal within uncertainties.  

\subsection{Vy\,1-2}  
This is a young PN with a complex morphology. \citet{akras:15} carried out a deep study of this object analysing its morphology, kinematics and chemistry. 
According to these authors the central star can be classified as a binary weak-emission-line star, $wels$, as it presents weak emission lines of \ion{C}{ii}, \ion{C}{iv} and \ion{O}{iii}.  Its effective temperature is in the range from $7.5 \times 10^4$ to $1.19 \times 10^5$ K. A possible very late thermal pulse (VLTP) could have occurred in the central star.
From their low-resolution spectrum \citet{akras:15} report a solar abundance of O, slightly enhanced N and depleted C. 

In this work we derived a heliocentric radial velocity of $-82.40 \pm 7.87$ km s$^{-1}$. Considering the distance reported by \citet{frew:16} (8.13 kpc) and by \citet{akras:15} (9.7 kpc),  and the heliocentric velocity, this PN could be a halo object as a height of $3.3 - 3.9$ kpc above the galactic plane is derived from its galactic latitude  $\rm{b} = +24^\circ$. However it shows an abundance $\rm{12+log(O/H) = 8.84}$ (see Table \ref{table:abundances}) which is too large for a halo PN. This PN shows log(N/O) abundance ratio of $-0.38$, $\rm{log(Ne/O) = -0.76}$, $\rm{log(S/O) = -1.82}$, $\rm{log(Ar/O) = -2.62}$, and $\rm{log(Cl/O) = -3.44}$. Compared to average values for disc PN abundances given by \citet{kingsburgh:94}, Vy\,1-2 abundances seem typical of a disc PN. 

For this object, we did not detect any Ar$^{+2}$ lines in our spectra.  
As we determined the total abundance by using \citet{kingsburgh:94} ICF, which takes into account the ionic abundances of Ar$^{+2}$, Ar$^{+3}$ and Ar$^{+4}$, our total Ar/H abundance should be considered as a lower limit, but closer to the real abundance; our value is consistent with those derived by \citet{wesson:05} and \citet{akras:15}.

We found a 12+log(K/H) ratio of $4.45_{-0.24}^{+0.16}$ which is lower than the solar value by about a factor of $\sim$5.1; however, the ICF used to correct the observed K$^{+3}$ abundance, which is based on the ionic fraction of Ar$^{+3}$, requires a reliable determination of Ar/H abundance \citep[and references therein]{amayo:20}. Therefore, our K/H value should be also considered as a lower limit of the real abundance.

\citet{wesson:05}   reported an ADF(O$^{+2}$) of 6.17 for this nebula. In this work we derived a value ADF(O$^{+2}$) of 5.34$_{-1.08}^{+1.27}$ which is consistent with their value.

The kinematics of  Vy\,1-2 is peculiar (Fig. \ref{fig:kinematics} middle-left). Because a central ring is the brightest structure \citep{akras:15}, the lines detected are mainly emitted in this zone. Expansion velocities provided by CELs show no gradient, actually the expansion velocities of ions in the inner zone (He$^{+2}$, Ar$^{+3}$, Ne$^{+2}$, O$^{+2}$) are larger than the velocities in the outer zone (N$^+$, O$^+$, S$^+$), which is opposite of what is expected in an expanding PN following a Hubble-law flow. Only the very inner ions (Ar$^{+4}$, Ne$^{+3}$) present low expansion velocities. Velocities from ORLs show  no systematic behaviour either, being some times lower and sometimes higher than CELs velocities. Therefore the inner zone seems to be accelerated  compare to the outer zone. In the literature few PNe present this peculiar kinematic behaviour; considering the 14 objects studied by \citet{pena:17}  and the 40 objects analysed by \citet{medina:06} only two of them,  PN BD+30 3639 and PN M\,1-32, show a similar behaviour with $\rm{V_{exp}}$ of [\ion{O}{iii}] and other high ionized species larger the $\rm{V_{exp}}$ of [\ion{N}{ii}], putting in evidence  high velocity gas in the central zone.  

\subsection{IC\,4997}

This is a very young and dense nebula, whose central star, classified as $wels$, has presented important changes and where the nebula is also changing, heating with time.
\citet{kostyakova:09} monitored the spectral evolution of this object for forty years. 
They claimed that the electron density increased from $4 \times 10^5$ to $2 \times 10^6$ cm$^{-3}$, and the electron temperature increased  from 12,000 K to 14,000 K in the period 1972 to 1992.  By considering the  \ion{He}{i}, [\ion{Ne}{iii}], and [\ion{O}{iii}] lines it is found that the nebular ionization degree has been growing with time. The central star seems to have increased its effective temperature from $37,000 - 40,000$ K to 47,000 K in the same period.

IC\,4997 was studied by \citet{arrieta:03}, who measured a FWZI(H$\alpha$) equivalent to 5,100 km s$^{-1}$ attributed to Raman scattering. This was also suggested by \citet{lee:00}. \citet{feibelman:92} reported a secondary emission component, displaced by $-69$ km s$^{-1}$ from the peak of the broad main body of H$\alpha$.

 \citet{flower:80} analysed this object from UV and optical spectra, founding abundances $\rm{12+log(O/H) = 8.04}$ and $\rm{12+log(C/H) = 7.65}$. 
 In this work we derived  $\rm{12+log(O/H) = 8.25_{-0.26}^{+0.34}}$ and an ADF(O$^{+2}$) of $4.87_{-2.71}^{+4.34}$. Then this object is O under-abundant relative to the Sun. 
 
 \citet{flower:80} attributed the low C and O abundances to the possibility that an important amount of these elements could be embedded in dust grains. In this work we find that S is also slightly under-abundant which might indicate that also S could be embedded in dust grains. 
 Even when we found abundances of Fe$^{+2}$ and K$^{+3}$, we could not use the previous ICF to determine total abundances because they are outside their validity range. 
 
The kinematics of IC\,4997 is very complex. No gradient of CELs or ORLs expansion velocities are found. The same as in the case of Vy\,2-2, the auroral and trans-auroral lines show lower expansion than the nebular lines of the same ion. Again this phenomenon can be attributed to the extreme density gradient in this nebula, which has a density of about 30,000 cm$^{-3}$ in the periphery and larger than about 10$^6$ cm$^{-3}$ in the inner zone. The auroral and trans-auroral lines would be mainly produced in a higher density zone, where nebular lines are undergoing collisional de-excitation due to they have lower critical densities.
 A more detailed discussion on this subject can be found in next subsection.
 
\subsection{Nebular and auroral line kinematics in Vy\,2-2 and IC\,4997}
 
As we mentioned in the previous subsections, in Vy\,2-2 and IC\,4997, auroral (trans-auroral) and nebular lines of [\ion{S}{ii}], [\ion{N}{ii}], [\ion{O}{iii}], [\ion{Ar}{iii}] and [\ion{Ar}{iv}] present different profiles and therefore different kinematic behaviour.

From very high-resolution spectra of PN NGC 6153, \citet{barlow:06} found that nebular and auroral lines of [\ion{O}{iii}] $\lambda$5007 and $\lambda$4363 present different velocity profiles. Since electronic temperature is estimated from this ratio, the authors argued that large variations in velocity of this ratio may be indicative of temperature fluctuations in the nebula. From photoionization models, \citet{zhang:08} explored if the differences between auroral and nebular lines profiles can be attributed to temperature or density variations in the plasma; he argues that if these effects are present in the nebulae, lines with different critical densities or excitation temperatures would show different profiles.
 
\begin{figure*}
	\includegraphics[width=\columnwidth]{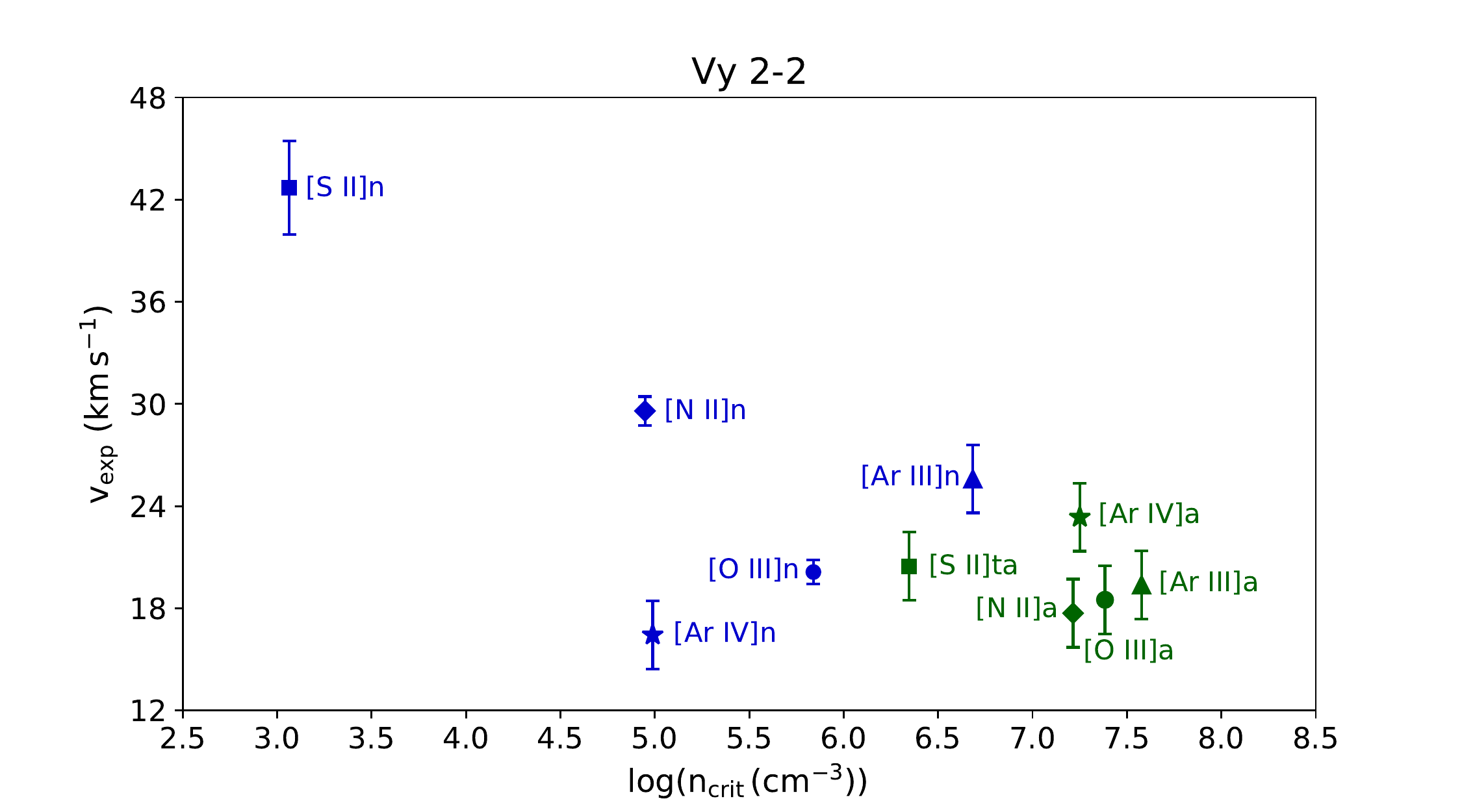}
	\includegraphics[width=\columnwidth]{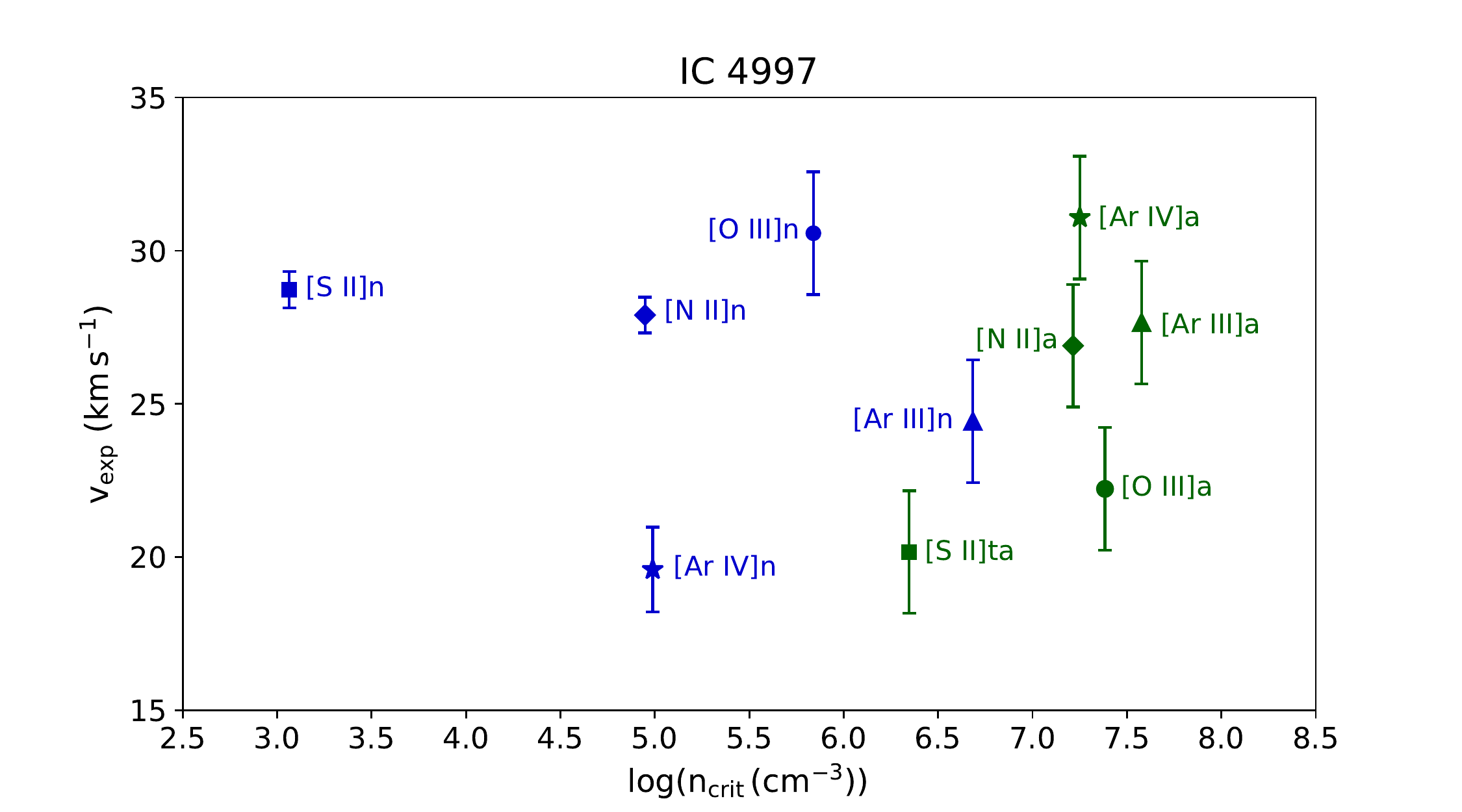}
    \caption{Expansion velocity vs. critical density (log scale) of nebular and auroral lines of [\ion{S}{ii}], [\ion{N}{ii}], [\ion{O}{iii}], [\ion{Ar}{iii}] and [\ion{Ar}{iv}] in Vy\,2-2 and IC\,4997. 
    Expansion velocities from nebular lines are in blue, expansion velocities from auroral are in green.}
    \label{fig:ncrit}
\end{figure*}

 We attribute the velocity differences between auroral and nebular lines in Vy\,2-2 and IC\,4997 to density variations and not to temperature variations because in both nebulae we found a strong density gradient increasing into their inner zones and the differences in velocity are strongly marked in lines which are not useful to determine electronic temperatures, such as [\ion{S}{ii}] $\lambda\lambda (6716+6731)/4068$ and [\ion{N}{ii}] $\lambda\lambda (6548+6584)/5755$ which in the present circumstances are sensitive to density.  To analyse better this phenomenon, we compare the critical densities and the expansion velocities of the auroral and nebular lines for these two objects (Fig. \ref{fig:ncrit}) in order to find a possible correlation between both parameters.
 
For Vy\,2-2 (Fig. \ref{fig:ncrit}, left) we found that auroral lines, which have critical densities larger than $10^6$ cm$^{-3}$, are concentrated in a zone with expansion velocities between 17 and 27 km s$^{-1}$. Nebular lines of [\ion{S}{ii}] and [\ion{N}{ii}] have larger expansion velocities than their auroral (trans-auroral) lines, by about 20 and 12 km\,s$^{-1}$ respectively. For the cases of [\ion{Ar}{iii}], [\ion{O}{iii}] and [\ion{Ar}{iv}] lines, the differences between auroral and nebular lines are not so considerable: for [\ion{Ar}{iii}] 
$\rm{V_{exp}}$ of the auroral line is 6 km\,s$^{-1}$ higher than the nebular value, however considering the errors of $\rm{V_{exp}}$ of [\ion{Ar}{iii}] lines the difference is just 2 km\,s$^{-1}$; in the case of [\ion{O}{iii}], auroral and nebular lines show the same expansion velocities of about 20 km\,s$^{-1}$; while for [\ion{Ar}{iv}], which shows the opposite behaviour, the velocity of the auroral line is 6 km\,s$^{-1}$ larger than the nebular but considering the errors the difference is just of 2 km\,s$^{-1}$. As we showed in Fig. \ref{fig:kinematics}, Vy\,2-2 presents a Hubble-type flow for expansion velocities of ions, therefore we can consider that auroral lines arises from inner zones of the nebula due to their lower expansion velocities, these inner zones have  higher density and the emission of lines with low critical densities is suppressed there. This effect can be found in the nebular-auroral line ratios of  [\ion{S}{ii}], [\ion{N}{ii}] and [\ion{Ar}{iv}], which are sensitive to density instead of  temperature (see Fig. \ref{fig:diagnostics}); nonetheless the density in inner zones of the nebula in not enough to overpass the critical densities of auroral lines of [\ion{Ar}{iii}] and [\ion{O}{iii}] ($\rm{n_{crit} > 10^{7.38}}$) and so the their line ratios are still sensitive to temperature.

 For IC\,4997 (Fig. \ref{fig:ncrit}, right), there is clear discrepancy between the velocities of the auroral and nebular lines of [\ion{S}{ii}], [\ion{O}{iii}] and [\ion{Ar}{iv}], the differences are of 11, 8 and 12 km\,s$^{-1}$, respectively; as in the case of Vy\,2-2, [\ion{Ar}{iv}] auroral line shows larger velocity than the nebular one. Auroral and nebular lines of [\ion{N}{ii}] and [\ion{Ar}{iii}] show the same expansion velocities (within the errors) of 27 and 25 km\,s$^{-1}$. As in Vy\,2-2, for the ions which show important kinematic discrepancies between auroral and nebular lines, their auroral-nebular ratio is not sensitive to temperature and it is sensitive to density. Given the kinematic complexity of the nebula, which does not follow a Hubble-type flow (Fig. \ref{fig:kinematics}), it is not possible to determine if the differences of velocities of the lines are indicative of inner or outer zones of the nebula, and they might be corresponding to complex structures within the nebula.

\defcitealias{milingo:10}{M10} 
\defcitealias{akras:15}{A15} 
\defcitealias{delgado:15}{DI15} 
\defcitealias{perinotto:04}{P04}

\begin{table*}
\centering

\caption{\bf Comparison of total abundances in 12+log(X/H) scale$^{a,b,c}$}
\begin{tabular}{lccccccccccccc}
\hline
 & Cn\,3-1 & Cn\,3-1 & Cn\,3-1 & Vy\,2-2 & Vy\,2-2 & Vy\,2-2 & Hu\,2-1 & Hu\,2-1 & Hu\,2-1 & Vy\,1-2 & Vy\,1-2 & Vy\,1-2 & IC\,4997$^c$ \\
 & This work & \citetalias{milingo:10} & \citetalias{wesson:05} & This work & \citetalias{perinotto:04} & \citetalias{wesson:05} & This work & \citetalias{delgado:15} & \citetalias{wesson:05} & This work & \citetalias{akras:15} & \citetalias{wesson:05} & This work \\ \hline \hline 
He/H & 10.73$\pm$0.02 & 10.74 & 10.65 & 11.06$\pm$0.01 & 10.88 & 11.03 & 11.03$\pm$0.02 & 10.91 & 10.90 & 11.08$\pm$0.03 & 11.04$\pm$0.03 & 11.03 & 11.16$\pm$0.01 \\
O/H & 8.60$_{-0.08}^{+0.12}$ & 8.59 & 8.63 & 8.14$_{-0.05}^{+0.12}$ & 8.16 & 7.98 & 8.46$_{-0.20}^{+0.38}$ & 8.30 & 8.51 & 8.84$_{-0.08}^{+0.10}$ & 8.66$\pm$0.04 & 8.70 & 8.25$_{-0.26}^{+0.34}$ \\
N/H & 7.82$\pm$0.05 & 7.89 & 7.87 & 7.74$_{-0.09}^{+0.13}$ & 7.42 & 7.83 & 7.87$_{-0.12}^{+0.11}$ & 7.73 & 7.74 & 8.46$_{-0.11}^{+0.10}$ & 8.13$\pm$0.05 & 8.06 & 7.63$_{-0.15}^{+0.20}$ \\
Ne/H & 6.79$\pm$0.16 & --- & --- & 7.58$_{-0.05}^{+0.11}$ & --- & 7.24 & 7.52$_{-0.14}^{+0.38}$ & 7.71 & 7.45 & 8.07$_{-0.10}^{+0.11}$ & --- & 8.00 & 7.60$_{-0.21}^{+0.26}$ \\
Ar/H & --- & 5.83 & 6.09 & 6.48$\pm$0.03 & 6.05 & 5.88 & 6.18$_{-0.23}^{+0.46}$ & 5.81 & 5.78 & 6.22$_{-0.09}^{+0.11}$ & 6.28$\pm$0.06 & 6.25 & 5.84$_{-0.20}^{+0.26}$ \\
S/H & 6.77$_{-0.05}^{+0.07}$ & 6.96 & --- & 7.31$\pm$0.05 & 6.38 & 6.65 & 6.37$_{-0.18}^{+0.36}$ & --- & 6.10 & 7.02$_{-0.10}^{+0.12}$ & 6.80$\pm$0.08 & 6.78 & 6.44$_{-0.18}^{+0.23}$ \\
Cl/H & 5.11$_{-0.06}^{+0.08}$ & 5.33 & --- & 4.97$_{-0.11}^{+0.23}$ & --- & --- & 4.70$_{-0.13}^{+0.27}$ & 4.68 & 4.59 & 5.38$\pm$0.18 & 5.22$\pm$0.12 & --- & 4.41$_{-0.18}^{+0.23}$ \\ \hline

N/O & $-0.78_{-0.09}^{+0.06}$ & $-$0.70 & $-$0.76 & $-0.41_{-0.08}^{+0.07}$ & $-$0.74 & $-$0.15 & $-0.60_{-0.18}^{+0.11}$ & $-$0.57 & $-$0.77 & $-0.38_{-0.15}^{+0.10}$ & $-$0.53$\pm$0.06 & $-$0.64 & $-0.62_{-0.20}^{+0.15}$ \\ 
Ne/O & $-1.82_{-0.11}^{+0.10}$ & --- & --- & $-0.56 \pm 0.02$ & --- & $-$0.74 & $-0.92 \pm 0.03$ & $-$0.59 & $-$1.06 & $-$0.76$\pm$0.03 & --- & $-$0.70 & $-0.66_{-0.08}^{+0.06}$ \\
Ar/O & --- & $-$2.76 & $-$2.54 & $-1.65_{-0.15}^{+0.06}$ & $-$2.11 & $-$2.10 & $-2.26_{-0.07}^{+0.11}$ & $-$2.49 & $-$2.73 & $-$2.62$\pm$0.03 & $-$2.38$\pm$0.07 & $-$2.45 & $-2.42_{-0.09}^{+0.07}$ \\
S/O & $-1.82_{-0.04}^{+0.06}$ & $-$1.63 & --- & $-0.85_{-0.13}^{+0.07}$ & $-$1.78 & $-$1.33 & $-2.08_{-0.05}^{+0.04}$ & --- & $-$2.41 & $-1.82_{-0.07}^{+0.06}$ & $-$1.86$\pm$0.09 & $-$1.92 & $-1.81 \pm 0.52$ \\
Cl/O & $-3.49\pm0.04$ & $-$3.26 & --- & $-3.16_{-0.08}^{+0.11}$ & --- & --- & $-3.75_{-0.12}^{+0.07}$ & $-$3.62 & $-$3.92 & $-$3.44$\pm$0.09 & $-$3.44$\pm$0.13 & --- & $-3.84_{-0.36}^{+0.25}$ \\ 
\hline
\multicolumn{14}{l}{$^a$ \citetalias{akras:15}:\citet{akras:15}, \citetalias{delgado:15}:\citet{delgado:15}, \citetalias{milingo:10}:\citet{milingo:10}, \citetalias{perinotto:04}:\citet{perinotto:04}, \citetalias{wesson:05}:\citet{wesson:05}. } \\
\multicolumn{14}{l}{$^b$ Abundances from \citet{flower:80} for O and C are presented in the text.} \\
\multicolumn{14}{l}{$^c$ Solar abundances by \citet{asplund:09} are: He/H=10.93, O/H=8.69, C/H=8.43, N/H=7.83, Ne/H= 7.93, Ar/H=6.40, S/H=7.12}\\ 

\end{tabular}
\label{tab:comparison_abundances}
\end{table*}

\section{Discussions and Conclusions}

From high-resolution spectra, physical conditions (temperature and density) and chemical abundances from CELs and ORLs have been determined for the PNe Cn\,3-1, Vy\,2-2, Hu\,2-1, Vy\,1-2 and IC\,4997, for which our main goal was to derive the ADFs(O$^{+2}$). The five objects are young and dense, show density gradients growing towards the inner zone and they have different ionization degree and different morphology. Abundance determinations of He, O, N, Ne, Ar, S, Cl have been obtained from CELs and our values are in general agreement with values previously reported in the literature.

Three of the objects, Cn\,3-1, Hu\,2-1, and Vy\,1-2, possess typical abundances of disc PNe, comparable to the solar values. Vy\,2-2 and IC\,4997 are low-abundance objects with 12+log(O/H) of about 8.2 or lower, which can be attributed to O depletion in dust grains or to central stars formed in a low abundance medium. The large values of Ne/O, Ar/O and S/O presented by both nebulae better indicate that O might be depleted in dust grains. We also were able to determine abundances of K/H for Vy\,1-2 and Fe/H for Vy\,2-2 and Hu\,2-1.

From comparison of chemical abundances from CELs and ORLs,
ADFs(O$^{+2}$) were calculated for Vy\,2-2, Hu\,2-1, Vy\,1-2 and IC\,4997, obtaining values of 4.30, 1.85, 5.34 and 4.87 respectively. For Vy\,2-2 and Hu\,2-1, our values resulted to be smaller than values previously reported in the literature. 
The difference between our values and the previously reported can be attributed to the selection of multiplets to determine the abundances for  \ion{O}{ii}. In our calculations we have determined this abundance  using the lines of multiplet V1 only, which is the strongest among \ion{O}{ii} multiplets; the lines from other multiplets (e.g., V2, V5, V10, V19, V20 and 3p-4f transitions) are much fainter than those from V1 and higher errors are expected in \ion{O}{ii} abundance. 
The Vy\,2-2, Hu\,2-1 and Vy\,1-2 \ion{O}{ii} abundances determined by \citet{wesson:05} include the contribution of other multiplets.
We found that  our \ion{O}{ii} V1 abundances for these objects are similar to those derived by \citet{wesson:05} for the same multiplet. However, when these authors include the contribution of the other multiplets, their \ion{O}{ii} abundances become higher leading to higher ADFs values. On the other side, \citet{wesson:05} did not report errors for the abundances of each multiplet and then it is not possible to know if those differences are within their respective error bars.

Considering the ADFs, the gas emitting ORLs is enriched by a large factor, relative to the gas emitting CELs. An important work to do is to estimate the amount of nebular mass contained in the gas emitting CELs and the gas emitting ORLs, to determine the real chemical abundances in the nebulae.

The temperatures derived from recombination lines of \ion{O}{ii} are lower than the values obtained from [\ion{O}{iii}] CELs. We did not find evidence of the existence of cold H-deficient He- and heavy elements-rich small inclusions embedded in the nebulae and emitting most of ORLs as proposed by \citet{liu:00,liu:06,liu:12}, among others, because according to their results, and also results from other authors \citep[e. g.,][]{tsamis:04,wesson:05}, such inclusions would have temperatures of about 1,000 K or even lower. We did not find such extremely low ORLs temperatures in the PNe of our sample, for all cases we found that these temperatures are above 6,000 K. 

The kinematics of the gas emitting CELs and ORLs was analysed for each nebula to study the possibility of different plasmas (with different physical conditions and spacial distribution) coexisting in the nebula and emitting the different lines. We compared in particular the expansion velocities given by CELs and ORLs emitted by O$^{+2}$. The results indicate that  such velocities are equal within uncertainties in Cn\,3-1, Vy\,2-2, Hu\,2-1 and Vy\,1-2, therefore from the kinematics point of view, there is no evidence for these lines being emitted in different zones of the nebula. But such velocities are very different in IC\,4997, where we find  $\rm{V_{exp}(\lambda4959) = 29.56 \pm 2}$ km s$^{-1}$ and $\rm{V_{exp}(\lambda4649) = 23.30 \pm 2}$ km s$^{-1}$, which might be implying that ORLs are being emitted in a different zone. According to Table \ref{table:recdensity-temperature}, the density and temperature given by O$^{+2}$ recombination lines in IC\,4997 are 5,000 cm$^{-3}$ and 7,700 K which are much lower than the values derived from CELs.

 As mentioned early for Vy\,2-2 and IC\,4997, we found the interesting fact that, in some cases, nebular and auroral (or trans-auroral) lines of the same ion (e.g., S$^+$, N$^{+}$, Ar$^{+2}$, Ar$^{+3}$, O$^{+2}$) present different expansion velocities. Auroral and trans-auroral lines (marked in green in Fig. \ref{fig:kinematics}) show, in general, lower $\rm{V_{exp}}$  which in the case of Vy\,2-2 (showing a Hubble velocity expansion law) might be indicating that auroral lines (sensitive to density and temperature) are being emitted in a denser and inner zone than the nebular lines. This might be a consequence of, in general, nebular lines have critical densities smaller than auroral lines, therefore nebular lines are undergoing collisional de-excitation and are not emitted in dense zones. The phenomenon is very complex in IC\,4997 where the auroral lines of S$^{+}$ and O$^{+2}$ present lower $\rm{V_{exp}}$ than nebular lines, and the opposite occurs in Ar$^{+3}$. Anyway, being Vy\,2-2 and IC\,4997 the objects showing the highest density differences between the inner and outer zones, this phenomenon seems more related to density than to temperature effects.

IC\,4997 is a very young PN, around a rapidly evolving hot post-AGB central star. This interesting object has been showing stellar and nebular evolution in short timescales, therefore it is necessary to keep tracking IC\,4997 and other similar objects in order to better understand their behaviour and fast evolution towards the PN stage.
 
A similar work aiming to determine ADFs in PNe with different characteristics is being carried out for a number of objects. It will be published in the future.

\section*{Acknowledgements}
We are indebted to Drs. Michael Richer, Christophe Morisset and Antonio Peimbert for interesting comments and suggestions along this research. We are grateful to Jos\'e N. Esp\'iritu for his help with \textsc{pyneb} and to Alexia Amayo for her help with the Monte Carlo error determinations. We thank an anonymous referee for his/her careful reading of the manuscript and his/her suggestions that help to improve this work. 
This work is based upon observations carried out at the Observatorio Astron\'omico Nacional at the Sierra San Pedro M\'artir (OAN-SPM), Baja California, M\'exico.
We thank the daytime and night support staff at the OAN-SPM for facilitating and helping to obtain our observations.  F.R.-E. acknowledges scholarship from CONACyT, M\'exico.
This work received partial support from DGAPA-PAPIIT  IN105020 and IN103519, UNAM. 
 \medskip
 
 {\bf Data Availability Statement:} The data underlying this article will be shared on reasonable request to the corresponding author.



\bibliographystyle{mnras}
\bibliography{adfs_hd_pne} 



\appendix

\section{Atomic data}

\begin{table}
\centering
	\caption{\bf Atomic parameters used in \textsc{pyneb} calculations}
\begin{tabular}{lcc} \hline
Ion & Transition probabilities & Collisional strenghts \\
\hline
N$^+$ & \citet{froese:04}  & \citet{tayal:11}\\
O$^+$ & \citet{froese:04} & \citet{kisielius:09}\\
O$^{+2}$ & \citet{froese:04} & \citet{storey:14} \\
         & \citet{storey:00} &  \\
Ne$^{+2}$ & \citet{galavis:97} & \citet{mclaughlin:00} \\
Ne$^{+3}$ & \citet{zeippen:82} & \citet{giles:81} \\
S$^+$ & \citet{podobedova:09} & \citet{tayal:10} \\
S$^{+2}$ & \citet{podobedova:09} & \citet{tayal:99} \\
Cl$^{+2}$ & \citet{mendoza:83} & \citet{butler:89} \\
Ar$^{+2}$ & \citet{mendoza:83} & \citet{galavis:95} \\
         & \citet{kaufman:86} &           \\
Ar$^{+3}$ & \citet{kaufman:86} & \citet{mendoza:82} \\
Ar$^{+4}$ & \citet{lajohn:93} & \citet{galavis:95} \\
          & \citet{mendoza:82} &           \\
          & \citet{kaufman:86} &           \\
K$^{+3}$ & \citet{mendoza:83} & \citet{galavis:95} \\
          & \citet{kaufman:86} &           \\
\hline

Ion & \multicolumn{2}{c}{Effective recombination coefficients} \\
\hline
H$^+$ & \multicolumn{2}{c}{\citet{storey:95}} \\
He$^+$ & \multicolumn{2}{c}{\citet{porter:12,porter:13}} \\
He$^{+2}$ & \multicolumn{2}{c}{\citet{storey:95} }  \\
N$^{+2}$ & \multicolumn{2}{c}{\citet{fang:11}}  \\
O$^{+2}$ & \multicolumn{2}{c}{\citet{storey:17} }  \\
C$^{+2}$ & \multicolumn{2}{c}{\citet{pequignot:91}}  \\
\hline
\end{tabular}
\label{tab:atomic-parameters}
\end{table}

\section{ionization Correction Factors}
\label{section:icfs_exp}

ICFs used for the total abundances calculation are listed next. 

\defcitealias{kingsburgh:94}{KB94} 
\defcitealias{delgado:14}{DI14} 

\begin{itemize}

\item $\rm{ \frac{He}{H} = \frac{He^{+}}{H^{+}} }$. If He$^{+2}$ is detected $\rm{ \frac{He}{H} = \frac{He^{+} + He^{+2}}{H^{+}} }$.

\item $\rm{ \frac{O}{H} = ICF(O) \times \frac{O^+ + O^{+2}}{H^+}}$. $\rm{ICF(O) = 1}$ if no He$^{+2}$ is detected. \\ Otherwise, $\rm{ICF(O)}$ is given by the equation (12) in \citet{delgado:14}. 

\item $\rm{ \frac{N}{H} = ICF(N) \times \frac{N^+}{H^+}}$. $\rm{ICF(N) = \frac{O}{O^+}}$ \citep{kingsburgh:94}.

\item $\rm{ \frac{Ar}{H} = ICF(Ar) \times \frac{Ar^{+2} + Ar^{+3} + Ar^{+4}}{H^{+}} }$. $\rm{ICF(Ar) = \frac{1}{1-N^+/N} }$ \citep{kingsburgh:94}.

\item $\rm{ \frac{Ne}{H} =  ICF(Ne) \times \frac{Ne^{+2}}{H^{+}} }$. $\rm{ICF(Ne) = \frac{O}{O^{+2}} }$ \citep{kingsburgh:94}.

\item $\rm{ \frac{S}{H} = ICF(S) \times \frac{S^{+} + S^{+2}}{H^{+}} }$. $\rm{ICF(S) = \left[ 1 - \left( 1 - \frac{O^+}{O} \right)^3 \right]^{-1/3}}$. \citep{kingsburgh:94}.

\item $\rm{ \frac{Cl}{H} = ICF(Cl) \times \frac{Cl^{+2}}{H^+}}$. $\rm{ICF(Cl) = \frac{S}{S^+2}}$ \citep{liu:00,wesson:05}.

\item $\rm{ \frac{K}{Ar} = \frac{K^{+3}}{Ar^{+3}} \times ICF(K)}$, $\rm{ICF(K)}$ is given by equation (11) in \citet{amayo:20}. 

\item $\rm{ \frac{Fe}{O} = \frac{Fe^{+2}}{O^{+}} \times ICF(Fe)}$, $\rm{ ICF(Fe) = 1.1 \left( \frac{O^+}{O^{+2}} \right)^{0.58} }$, \citep{rodriguez:05}.



\end{itemize}



\bsp	
\label{lastpage}
\end{document}